\documentclass[a4paper,12pt]{article}
\usepackage{graphicx, rotating}

\ifx\pdfoutput\undefined
\usepackage[dvips,bookmarks]{hyperref}	
\else
\usepackage{hyperref}	
\fi
\hypersetup{colorlinks,bookmarksopen,bookmarksnumbered,citecolor=verdes,
linkcolor=blus,pdfstartview=FitH,urlcolor=rossos}
\def\hhref#1{\href{http://arxiv.org/abs/#1}{#1}} 

\usepackage{multicol}
\usepackage{color}
\definecolor{rosso}{cmyk}{0,1,1,0.4}
\definecolor{rossos}{cmyk}{0,1,1,0.55}
\definecolor{rossoc}{cmyk}{0,1,1,0.2}
\definecolor{blu}{cmyk}{1,1,0,0.3}
\definecolor{blus}{cmyk}{1,1,0,0.6}
\definecolor{bluc}{cmyk}{1,1,0,0.1}
\definecolor{verde}{cmyk}{0.92,0,0.59,0.25}
\definecolor{verdec}{cmyk}{0.92,0,0.59,0.15}
\definecolor{verdes}{cmyk}{0.92,0,0.59,0.4}

\font\tenrsfs=rsfs10 at 12pt
\font\sevenrsfs=rsfs7
\font\fiversfs=rsfs5
\newfam\rsfsfam
\textfont\rsfsfam=\tenrsfs
\scriptfont\rsfsfam=\sevenrsfs
\scriptscriptfont\rsfsfam=\fiversfs
\def\mathscr#1{{\fam\rsfsfam\relax#1}}

\newcommand{\fig}[1]{~\ref{fig:#1}}
\newcommand{\eq}[1]{~{\rm (\ref{eq:#1})}}

\newcommand{\GeV}{\,{\rm GeV}}
\newcommand{\TeV}{\,{\rm TeV}}
\newcommand{\eV}{\,{\rm eV}}

\def\circa#1{\,\raise.3ex\hbox{$#1$\kern-.75em\lower1ex\hbox{$\sim$}}\,}

\newcommand{\DM}{{\rm DM}}

\newcommand{\PR}{Phys. Rev.}

\newcommand{\beq}{\begin{equation}}
\newcommand{\eeq}{\end{equation}}
\newcommand{\MeV}{\,{\rm MeV}}
\def\circa#1{\,\raise.3ex\hbox{$#1$\kern-.75em\lower1ex\hbox{$\sim$}}\,}
\makeatletter

\def\art{\@ifnextchar[{\eart}{\oart}}
\def\eart[#1]#2#3#4#5#6{{\rm #2}, {#3 #4} {\rm (#6) #5} [{\hhref{#1}}]}
\def\hepart[#1]#2{{\rm #2, \hhref{#1}}}
\newcommand{\oart}[5]{{\rm #1}, {#2 #3} {\rm (#5) #4}}

\newcounter{alphaequation}[equation]
\def\thealphaequation{\theequation\hbox to
0.6em{\hfil\alph{alphaequation}\hfil}}
\def\eqnsystem#1{
\def\@eqnnum{{\rm (\thealphaequation)}}
\def\@@eqncr{\let\@tempa\relax \ifcase\@eqcnt \def\@tempa{& & &} \or
  \def\@tempa{& &}\or \def\@tempa{&}\fi\@tempa
  \if@eqnsw\@eqnnum\refstepcounter{alphaequation}\fi
\global\@eqnswtrue\global\@eqcnt=0\cr}
\refstepcounter{equation} \let\@currentlabel\theequation \def\@tempb{#1}
\ifx\@tempb\empty\else\label{#1}\fi
\refstepcounter{alphaequation}
\let\@currentlabel\thealphaequation
\global\@eqnswtrue\global\@eqcnt=0 \tabskip\@centering\let\\=\@eqncr
$$\halign to \displaywidth\bgroup \@eqnsel\hskip\@centering
$\displaystyle\tabskip\z@{##}$&\global\@eqcnt\@ne
\hskip2\arraycolsep\hfil${##}$\hfil& \global\@eqcnt\tw@\hskip2\arraycolsep
$\displaystyle\tabskip\z@{##}$\hfil
\tabskip\@centering&\llap{##}\tabskip\z@\cr}
\def\endeqnsystem{\@@eqncr\egroup$$\global\@ignoretrue} \makeatother

\newcommand{\SU}{\rm SU}

\begin{document}
\begin{center}
{IFUP--TH/2007-12}
{ \hfill SACLAY--T07/052}
\color{black}
\vspace{0.3cm}

{\Huge\bf\color{rossos} Cosmology and Astrophysics of\\[2mm]
Minimal Dark Matter}
\medskip
\bigskip\color{black}\vspace{0.6cm}

{
{\large\bf Marco Cirelli}$^a$,
{\large\bf Alessandro Strumia}$^b$,
{\large\bf Matteo Tamburini}$^c$
}
\\[7mm]
{\it $^a$ Service de Physique Th\'eorique, CEA-Saclay, France and INFN, Italy}\\[3mm]
{\it $^b$ Dipartimento di Fisica dell'Universit{\`a} di Pisa and INFN, Italia}\\[3mm]
{\it $^c$ Dipartimento di Fisica dell'Universit{\`a} di Pisa}
\end{center}

\bigskip\bigskip

\centerline{\large\bf\color{blus} Abstract}
\begin{quote}
We consider DM that only couples to SM gauge bosons and
fills one gauge multiplet, 
e.g.\ a fermion 5-plet (which is automatically stable), or
a wino-like 3-plet.
We revisit the computation of the cosmological relic abundance 
including non-perturbative corrections.
The predicted mass of e.g.\ the 5-plet increases from $4.4\TeV$ to $10\TeV$,
and indirect detection rates are enhanced by 2 orders of magnitude.
Next, we show that due to the quasi-degeneracy among
neutral and charged components of the DM multiplet,
a significant fraction of DM with energy $E\circa{>} 10^{17}\eV$
(possibly present among ultra-high energy cosmic rays)
can cross the Earth exiting in the charged state and may in principle be detected in neutrino telescopes.
\color{black}
\end{quote}


\tableofcontents

\newpage

\section{Introduction}


%

Observations and experiments tell that:
\begin{itemize}
\item[i)] Dark Matter (DM) exists~\cite{DMreview}, with abundance $\Omega_{\rm DM} h^2 = 0.110\pm 0.005$~\cite{cosmonu};
\item[ii)] DM is cosmologically stable;
it has no electric charge~\cite{Gould}, no strong interactions~\cite{strong} and the DM coupling to the
$Z$ boson is smaller than $\sim 10^{-3} g_2$~\cite{Zcoupling}.
\end{itemize}
i) suggests that DM might be a weak-scale particle:
indeed $\Omega_{\rm DM}\sim 1$ is the typical
freeze-out abundance of a typical particle with weak-scale
mass $M$ and coupling $g$: $M/g \sim \sqrt{T_0 M_{\rm Pl}}\sim \TeV$.
ii) can be satisfied in appropriate models, 
although these properties are not  typical weak scale particles.

We here want to study Minimal DM (MDM) models: we add to the Standard Model (SM) 
one new multiplet, that only has gauge interactions and
a $\SU(2)_L$-invariant mass term $M$.
In this context,
ii) singles out 
scalars or fermions with zero hypercharge that fill a $\SU(2)_L$ representation
with odd dimension: $n=3,5,\ldots$~\cite{MDM}.
The neutral $\DM^0$ component is then lighter than the charged $\DM^\pm$ components,
by only $166\MeV$.
$\DM^0$ interacts with ordinary matter via one-loop exchange of $W^\pm$:
future direct DM searches will hopefully reach the sensitivity needed to detect it.

The fermion 5-plet is particularly interesting: 
all interactions other than the 
gauge interactions are incompatible with gauge and Lorentz symmetries~\cite{MDM}, so that it is automatically stable thanks to an accidental symmetry, like the proton in the SM. Other candidates that we will consider lack this feature (i.e. some extra symmetry is responsible for their stability) but are also interesting.
The fermion 3-plet is well known from supersymmetry with matter-parity, as `pure wino' DM;  the fermion doublet case as `pure higgsino' DM.
We will also study scalar MDM candidates.
In general, this scenario has one free parameter, the DM mass $M$, already fixed by  $\Omega_{\rm DM}$, such that all DM signals are univocally predicted. We here address the following issues.



%

\bigskip

First, we complete the computation of the cosmological abundance performed in~\cite{MDM}
taking into account $p$-wave annihilations and renormalization of the gauge couplings
up to the DM scale, and, more importantly, we include non-perturbative Sommerfeld corrections:
the DM wave-functions get distorted by Coulomb-like forces mediated by SM vectors.
Their relevance was pointed out in~\cite{HisanoCosmo},
in the case of gluino and wino as lightest supersymmetric particle.
In section~\ref{short} we discuss the basic physics of 
non-perturbative Sommerfeld corrections, and 
in section~\ref{OmegaDM} we show our results for the cosmological DM abundance.

Next, in section~\ref{indirect} we compute
the $\DM^0 \DM^0$ annihilation rates relevant for indirect MDM signals,
including the sizable enhancement due to Sommerfeld corrections.

Finally, in section~\ref{Astro} we study the possibility that the ultra-high-energy cosmic rays (UHE CR)
contain some DM particle (although this looks difficult within the standard CR acceleration mechanism),  showing that MDM candidates can cross the earth arriving to the detector
in the charged state. We discuss how it could manifest in detectors such as neutrino telescopes.

Section~\ref{concl} presents our conclusions.

\section{Non perturbative Sommerfeld corrections}\label{short}
If the DM mass is $M \approx M_Z$
and if the DM coupling is $g\approx g_2$,
DM DM annihilations into SM vectors have a too large cross section,
and consequently give a too low freeze-out DM abundance.
Since $\Omega_{\rm DM}\circa{\propto} M$ for $M> M_Z$, 
the observed DM abundance is obtained for a value of $M$ sufficiently larger than $M_Z$.\footnote{
Another solution exists for $M$ sufficiently lower than $M_Z$, 
such that annihilation into vectors are kinematically suppressed.
In the minimal scenario that we consider,
this solution is excluded by LEP2 and other collider data, but in general
it is still allowed in some part of the parameter range of
non-minimal scenarios~\cite{InertHiggs}.

A DM mass $M \approx M_Z$ can of course also be obtained by reducing $g$,
i.e.\ by assuming that DM is any $\SU(2)_L$ multiplet appropriately mixed with
a singlet.}
Since this value turns out to be in the TeV or multi-TeV range,
we can ignore SM particle masses when computing the annihilation rates.
Using SU(2) algebra, ref.~\cite{MDM} obtained a single expression 
for the cosmological abundance of any MDM candidate, taking into account all (co-)annihilations
in  $s$-wave approximation.
We here add $p$-wave annihilations and renormalization of the gauge couplings up to the DM mass $M$: each one of these effects gives a ${\cal O}(5\%)$ correction to $\Omega_{\rm DM}$.
Precise formul\ae{} are given in appendix~\ref{Pert}.

\medskip

Furthermore, we take into account non-perturbative electroweak Sommerfeld corrections.
Their relevance was pointed out in~\cite{HisanoCosmo}  and might look surprising,
so we start with a general semi-quantitative discussion before
proceeding with the detailed computations.
Scatterings among charged particles due to point-like interactions
are distorted by the Coulomb force, when the kinetic energy is
low enough that the electrostatic potential energy is relevant.
This leads e.g.\ to significant enhancements of the $\mu^-\mu^+$
annihilation cross section (attractive force) or to significant
suppressions of various nuclear processes (repulsive force).
These effects can be computed with a formalism developed by Sommerfeld~\cite{Sommerfeld}.
In the language of Feynman graphs, these effects are described by multi-loop photon ladder diagrams.
Perturbative computations would include only the first few diagrams:
the Sommerfeld enhancement is non-perturbative because 
a resummation of all ladder diagrams is needed.\footnote{
Sometimes in the literature `non-perturbative' is used with a different meaning:
to denote  effects that vanish in the perturbative limit because suppressed
e.g.\ by $e^{-1/\alpha}$ factors.}
The generalization of the Sommerfeld formalism to the case of DM DM annihilations,
that involves non abelian massive vectors,
was presented in~\cite{Hisano}.

\smallskip

Let us first discuss Sommerfeld corrections due to one
abelian vector with mass $M_V$ and gauge coupling $\alpha$: 
this case already contains the relevant physics
and can be analyzed in a simpler way.
Cross sections at  low energies are dominated by $s$-wave scattering,
with $p$-wave giving corrections of relative order $T/M$.
Since the DM DM annihilation is local (i.e.\ it occurs when the distance between the two DM particles
is $r\simeq 0$), the non perturbative correction is given by
$R=|\psi(\infty)/\psi(0)|^2$, where $\psi(r)$ is the (reduced) $s$-wave-function for the two-body
DM DM state with energy $K$, that in the non-relativistic limit 
satisfies
the  Schr\"odinger equation
\beq \label{eq:S}
-\frac{1}{M}\frac{d^2 \psi}{dr^2}+ V \cdot \psi= K \psi\qquad
V=\pm \frac{\alpha}{r} e^{-M_V r}
\eeq
with outgoing boundary condition $\psi'(\infty)/\psi(\infty) \simeq i M\beta$.
Here $K = M\beta^2$ is the kinetic energy of the two DM particles in the center-of-mass
frame, where each DM particle  has velocity $\beta$.
The $-$ sign corresponds to an attractive potential and the $+$ sign to a repulsive potential.
We define $\epsilon \equiv M_V/M$, the adimensional ratio between the vector mass and the DM mass.

\smallskip

Eq.\eq{S} is the prototype of the equations that 
describe corrections to  $\DM^0\DM^0$ annihilations mediated by
a $Z$ boson, or $\DM^+\DM^-$ or $\DM^+\DM^+$ co-annihilations mediated
by a $\gamma$ or $W$ or $Z$,
or corrections mediated by a gluon
(in models where coannihilations with and between colored particles are relevant).
In the three cases, the coupling $\alpha$ that appears in eq.\eq{S} would be
$\alpha_2$, $\alpha_{\rm em}$ and $\alpha_{\rm s}$ respectively.
In the non abelian case, eq.\eq{S} becomes a matrix equation and $V$ is the sum
of the various contributions mediated by all SM vectors.
Higgs exchange can also be relevant, in models where DM sizably couples to the Higgs.

\begin{figure}[t]
\begin{center}
\includegraphics[width=\textwidth]{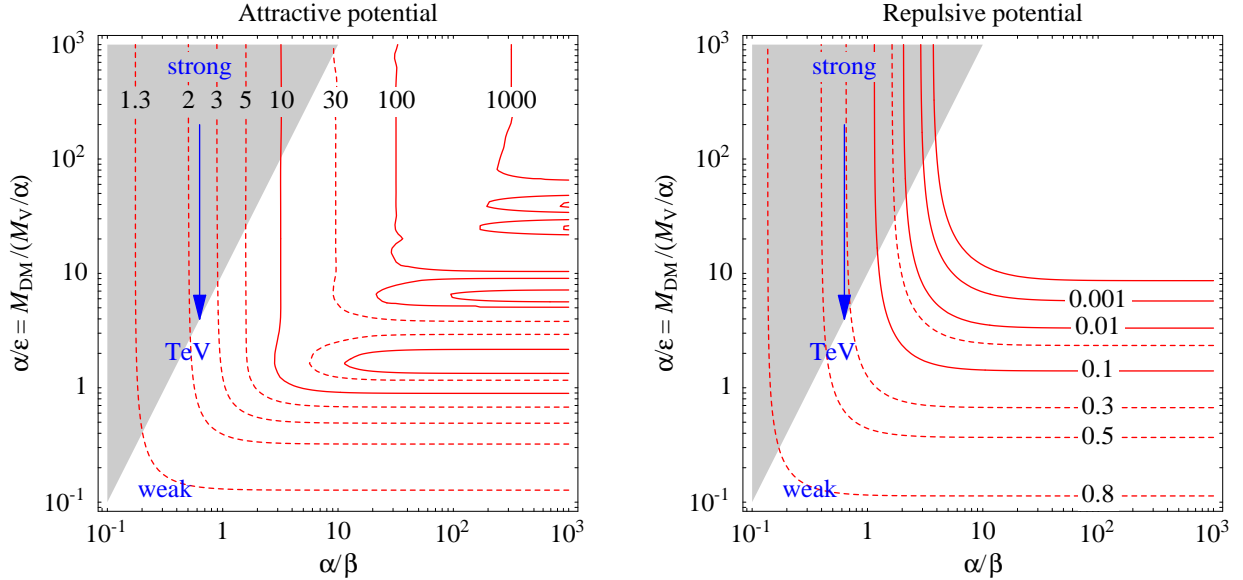}
\caption{\label{fig:nonpertd}\em Iso-contours of the non-perturbative Sommerfeld
correction to the DM DM annihilation. Here $\alpha$ is the coupling constant,
$\beta$ is the DM velocity, $\epsilon$ is the ratio between the vector mass and the DM mass.
The labels indicate where some classes of DM candidates lie in this plane:
`weak' indicates weak-scale DM particles, `TeV' indicates DM with multi-TeV mass,
and `strong' indicates strongly-interacting particles that in some models give dominant co-annihilations.
Within the shaded region thermal masses dominate over masses, 
effectively shifting the value of $\alpha/\epsilon$ as indicated by the arrow.
}
\end{center}
\end{figure}


\smallskip

For $\epsilon=0$ (massless vector) the Schr\"odinger equation has the same form
as the one that describes e.g.\ the hydrogen atom, and it can be analytically solved:
the Sommerfeld factor $R$ that multiplies the perturbative cross section is
\beq  \label{eq:R0}
R = \frac{-\pi x}{1-e^{\pi x}} \qquad x = \pm \frac{\alpha}{\beta}\qquad
\hbox{(for $M_V=0$)}\eeq
This shows that $R$ sizably differs from 1 at $\beta\circa{<}\pi\alpha$.
DM DM annihilations into SM particles freeze-out when the temperature $T$ cools down below
 $T/M\sim 1/\ln(M_{\rm Pl}/M) \sim 1/26$.
This happens to be numerically comparable to the SM gauge couplings, i.e.\ $\alpha_2 \approx 1/30$.
Consequently 
DM freeze-out occurs when $\beta\sim 0.2$ i.e.\ $\pi x \sim 1$:
the Sommerfeld correction is significant, $R\sim {\cal O}(1)$.

\medskip

Of course we must take into account that
the relevant $W,Z$ vectors are massive: since $R$
must now be computed numerically, it is convenient to
first identify on which parameters $R$ depends.
We notice that $R$ only depends on the two ratios $\alpha/\beta$ and $\alpha/\epsilon$,
so that $R$ can be plotted on a plane.
This can be proofed noticing that $R$ is adimensional and that physics is invariant under
$r\to \lambda r$, $M\to M/\lambda$, $M_V\to M_V/\lambda$.
Fig.\fig{nonpertd} shows iso-contours of $R$ as function of $\alpha/\beta$ and $\alpha/\epsilon$.
We can distinguish various regions:

\begin{itemize}
\item[a)] Non-perturbative effects are negligible (i.e.\ $|R-1|\ll1$) if $\pi  \alpha \ll \beta$ and/or if
$\alpha \ll \epsilon$.
We have seen that $\pi\alpha \sim \beta$, so non-perturbative effects  are relevant 
when $M \circa{>} M_V/\alpha$,
where $M_V$ are the SM vector masses and $\alpha$ are their gauge couplings~\cite{Hisano}.

\item[b)] If $\beta\circa{<}\epsilon\circa{<}\alpha$ (`lower triangle region')
$\beta$ is so low that its value is no longer relevant: $R$ depends only on $\alpha/\epsilon$
showing, in the attractive case, a series of resonances.
Indeed, by increasing $\alpha/\epsilon$ the potential develops more and more bound states:
resonant enhancement happens when a bound state first appears 
with energy  $E_B$ just below zero:
in this case, at low energy $R$ depends on $K$
as dictated by a Breit-Wigner factor  $1/(K-E_B)$.

\item[c)] If $\epsilon\circa{<}\beta\circa{<} \alpha $ (`upper triangle region')
non-perturbative effects depend almost only on $\alpha/\beta$:
vector masses have a minor effect and $R$ reduces to eq.\eq{R0}.
Its value does not depend on whether a loosely bound resonance is present or not.


\end{itemize}
When DM DM bound states are relevant, the Sommerfeld correction might not encode the full dynamics:
one should compute the density and life-time of DM DM bound states.

\medskip

We include one more effect, not discussed in~\cite{HisanoCosmo}.
At finite temperature $T$, the $\gamma,Z,W$ masses 
and the 
mass splitting $\Delta M$ between the components of the DM multiplet
are different than at zero temperature.
First, because these masses are proportional to
the SU(2)-breaking Higgs vev $v$, that depends on $T$
and vanishes at $T>T_c$ when SU(2) invariance gets restored by thermal effects,
via a second-order phase transition.
This effect can be roughly approximated as~\cite{TFT}
\beq v(T) = v \,{\rm Re} (1 - T^2/T_c^2)^{1/2} .\eeq
The critical temperature $T_c$ depends on the unknown Higgs mass
($T_c\simeq m_H$ for $m_H \gg v$): we here assume $T_c = 200\GeV$.

Second, the squared masses of all SM vectors $W,Z,\gamma,g$ get an extra contribution
known as thermal mass.
More precisely, even in the non abelian case, the Coulomb force gets screened
by the thermal plasma: this can be described by a vector Debye mass, 
equal to~\cite{vectors}
\beq \label{eq:mT}
m_{{\rm U(1)}}^2 = \frac{11}{6}g_Y^2 T^2,\qquad 
m_{\SU(2)}^2 = \frac{11}{6}g_2^2T^2,\qquad 
m_{\rm SU(3)}^2 = 2 g_3^2 T^2 \eeq
in the SM at $T \gg M_{W,Z}$.
We approximate vector masses by summing the squared masses
generated from $v(T)$ with the SU(2)-invariant Debye squared masses of eq.\eq{mT}.

\medskip

Let us now estimate the relevance of these thermal effects.
For the $W$ and $Z$ bosons, thermal masses dominate at $T\circa{>}v$,
and for the photon and the gluons  thermal masses dominate always.
At temperature $T$ DM particles have a typical energy $K\approx T$, 
so that thermal masses act like a contribution to $\epsilon$ of order
$\epsilon_T  \sim \sqrt{4\pi \alpha} \beta^2$. 
When this thermal effect is relevant,  
in fig.\fig{nonpertd} one has to shift $\alpha/\epsilon$ vertically down to
the diagonal of the shaded triangle:
we see that (depending on the precise value of $\alpha/\beta$) this shift
has a mild or negligible
effect on $R$, since thermal effects are relevant when
$R$ does not have a significant dependence on $\epsilon$.

\subsection{Computing Sommerfeld corrections}\label{somm}
Following~\cite{HisanoCosmo,Hisano}, we now 
give a operative summary about how to compute Sommerfeld corrections to $s$-wave
DM DM annihilation rates.
The set of two-body $\DM_i\, \DM_j$ states that mix among them is labeled as $i,j=\{1,\ldots N\}$.
For example, when studying  the wino triplet one encounters the 
$\{{\rm DM}^-{\rm DM}^+, {\rm DM}^0{\rm DM}^0\}$ system.
The dynamics is encoded in the $N\times N$ potential matrix $V$ and in
another $N\times N$  matrix $\Gamma$,
that describe the tree-level (co)-annihilation rates.
The strategy is the same as in the abelian case of eq.\eq{S}, plus a careful book-keeping of indices.
For each $j$ one solves the following set of $N$ coupled differential equations 
for $\psi_i^{(j)}(r)$:
\beq -\frac{1}{M} \frac{\partial^2  \psi_i^{(j)} }{\partial r^2} + \sum_{i'=1}^N
V_{ii'} \psi_{i'}^{(j)} = K \psi_i^{(j)} \eeq
with boundary conditions:
\beq
\psi_i^{(j)}(0) = \delta_{ij},\qquad \frac{\partial \psi^{(j)}_i(\infty)}{\partial r} = \sqrt{ M(K-V({\infty})_{ii})} \psi_i^{(j)}(\infty)
\eeq
chosen such that each wave has the same normalization at $r=0$,
where the annihilation occurs as described by the $\Gamma$ matrices.
The annihilation cross section $\sigma_i$ with given initial state $i$ at $r=\infty$
is obtained by 
factorizing out the oscillating phase,
$A_{ij} \equiv \lim_{r\to \infty} \psi_i^{(j)}(r)/e^{i\Re \sqrt{M (K-V(\infty))} r}$, and contracting with the annihilation matrix:
\beq \sigma_i  =c_i
(A\cdot \Gamma\cdot A^\dagger)_{ii}\eeq
The factor $c_i$ is given by
$c_i=2$ if the initial state $i$ has two equal DM particles
(e.g.\ as in the ${\rm DM}^0{\rm DM}^0$ state)
and $c_i=1$ otherwise (e.g.\ as in the ${\rm DM}^-{\rm DM}^+$ state).
This formalism automatically takes into account that some states
cannot exist as free particles at $r \gg 1/M_W$, when the kinetic energy $K$ is below their mass.

\bigskip

We next need to compute the Coulomb-like potential matrices $V$ and
the tree-level annihilation matrices $\Gamma$ for all DM components.
They split into sub-systems with given values of the $L$ (orbital angular momentum),
$S$ (total spin) and
$Q$ (total electric charge) quantum numbers.
Recalling that we neglect SM particle masses and that we consider $s$-wave 
(i.e.\ $L=0$)
tree-level annihilations into two SM particles,
annihilations obey the following selection rules:
\begin{itemize}
\item[0)]  two-body DM~DM states with spin $S=0$ can annihilate into two SM vectors,
that can have electric charge $Q=0,\pm1, \pm2$.

\item[1)] two-body DM~DM states with spin $S=1$
can annihilate into two SM fermions and into two SM Higgses,
that can have electric charge $Q=0,\pm1$.\footnote{Non-perturbative corrections are mostly relevant when 
$\SU(2)_L$ is unbroken: in this limit
one could perform a simplified analysis by replacing the 
electric charge quantum number with the $\SU(2)_L$ isospin quantum number $I$.
Then, the only DM DM states that can annihilate  into SM particles 
are those with $S=0$ and $I=1$ and $5$
(that annihilate into SM vectors; $I=1$ indicates the singlet) and the state with $S=1$ and  $I=3$ (that annihilates into SM fermions and Higgses.).}

\end{itemize}
The initial DM~DM state with $S=1$ of case 1) does not exist if DM is a scalar.\footnote{
These selection rules also apply to DM$^0$ DM$^0$ annihilations
relevant for indirect astrophysical DM signals, and reproduce well known results.
E.g.\ if DM$^0$ is a Majorana fermion the spin-statistics relation forbids $S=1$,
so that annihilations into vectors are not possible.
A non vanishing amplitude for DM$^0$ DM$^0\to f\bar{f}$, proportional to $m_f$,
is allowed when one takes into account the small mass $m_f$ of the SM fermions $f$.}

\medskip 

We now give explicit expressions for the annihilation matrices $\Gamma_{ii',jj'}$
and potential matrices $V_{ii',jj'}$.
Their off-diagonal entries do not have an intuitive physical meaning, and
are precisely defined in terms of the imaginary part and of the real part
of the two-body propagator $i i' \to jj'$, where the indices $ii'$ 
denote the two DM component in the initial state,
and $jj'$ denote the two DM components in the final state.

For MDM scalars with $Y=0$ one has (all states have $S=0$)
\begin{eqnsystem}{sys:GammaV}
 \Gamma_{ii', jj'} &=& \frac{N_{ii'}N_{jj'}}{8\pi n g_{\rm DM} M^2}\sum_{A,B}\{T^A, T^B\}_{ii'}\{T^A,T^B\}_{jj'}\\
   V_{ii'; jj'} (r)&=&   (M_i + M_j-M)\delta_{ii'}\delta_{jj'}+
  N_{ii'}N_{jj'} \sum_{AB} K_{AB}  (T^A_{ij} T^B_{i'j'}  + T^A_{ij'} T^B_{i'j})
  \label{eq:V0}
\frac{e^{-m_A r}}{r}
\end{eqnsystem}
where $N_{ij} = 1$ if $i\neq j$ and $1/\sqrt{2}$ if $i=j$.
The indices $A,B$ run over the SM vectors $\{\gamma, Z,W^+,W^-\}$ with generators $T^A$
in the DM representation.
The gauge couplings are included in the generators, such that the gauge-covariant derivative is
$D_\mu  = \partial_\mu + i A_\mu^A T^A$.
We emphasize that the (off-diagonal) entries of $T^\pm$ are 
defined up to an arbitrary phase normalization for each component of the multiplet: 
one needs to choose any convention such that the SU(2)-invariant DM mass term
has the same sign for all DM components.
The matrices $K_{AB}$ and $C_{AB}$ respectively describe SM vector propagation and
the gauge content of all other SM particles, and are given by:
{\small $$
K_{AB} = \pmatrix{ 
 1&0&0&0\cr
0&1&0&0\cr
0&0&0&1\cr
0&0&1&0},\qquad
C_{AB}= \frac{25 g_2^2}{4}
\pmatrix{s^2 & sc & 0 &0\cr
sc & c &0&0\cr
0&0&0&1\cr
0&0&1&0} + \frac{41}{4}g_Y^2
\pmatrix{c^2 
& -sc & 0 &0\cr
-sc & s^2 &0&0\cr
0&0&0&0\cr
0&0&0&0}.$$}

For MDM fermions with $Y=0$ one has
\begin{eqnsystem}{sys:GammaV}
 \Gamma_{ii', jj'}^{S=0} &=& \frac{N_{ii'}N_{jj'}}{16\pi n g_{\rm DM} M^2}\sum_{A,B}\{T^A, T^B\}_{ii'}\{T^A,T^B\}_{jj'}\\
  \Gamma_{ii', jj'}^{S=1} &=& \frac{N_{ii'}N_{jj'}}{8\pi n g_{\rm DM} M^2} \sum_{A,B} T^A_{ii'} T^B_{jj'} C_{AB}\end{eqnsystem}
$\Gamma^{S=1}$ is already summed over the 3 spin components. 
The potentials $V$ remain the same as in the scalar case of eq.\eq{V0},
up to an extra $-$ sign that 
must be taken into account when computing
elements of $V$ mediated by the $W$, such as $V_{0-,0-} =-V_{0-,-0}$,
because an extra $-$ sign appears in the definition of two body fermion states:
$|\DM^0\DM^-\rangle = - |\DM^-\DM^0\rangle$.
Furthermore, Fermi statistics forbids the existence of
some states, such as $|\DM^0\DM^0\rangle$ with $S=1$.

\medskip

Similar formul\ae{} apply to MDM candidates with  $Y\neq 0$,
after taking into account that $\DM^0$ now becomes a complex particle 
(its normalization factor changes from $N_{00}=1/2$ to $N_{0\bar 0}=1$)
and that one is only interested in DM $\overline{\rm DM}$ annihilations and two body states.
Summing $\Gamma_{ii',ii'}$ over all components one reproduces
the total co-annihilation rates of~\cite{MDM}, here reported and extended in appendix~\ref{Pert}.
Explicit expressions for $V$ and $\Gamma$ are given in the next sections.

\section{Cosmological abundances}\label{OmegaDM}
To include non-perturbative corrections a dedicated lengthy analysis is needed for each MDM candidate.

\begin{figure}[t]
\begin{center}
\includegraphics[width=0.45\textwidth]{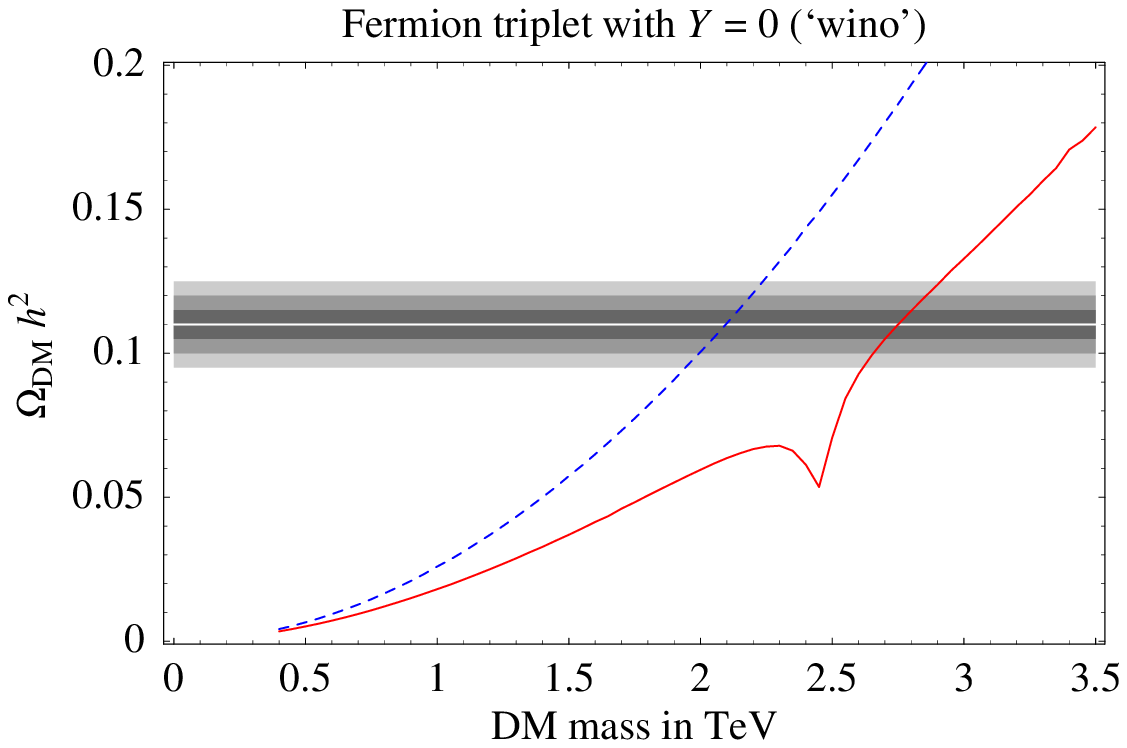}
\includegraphics[width=0.45\textwidth]{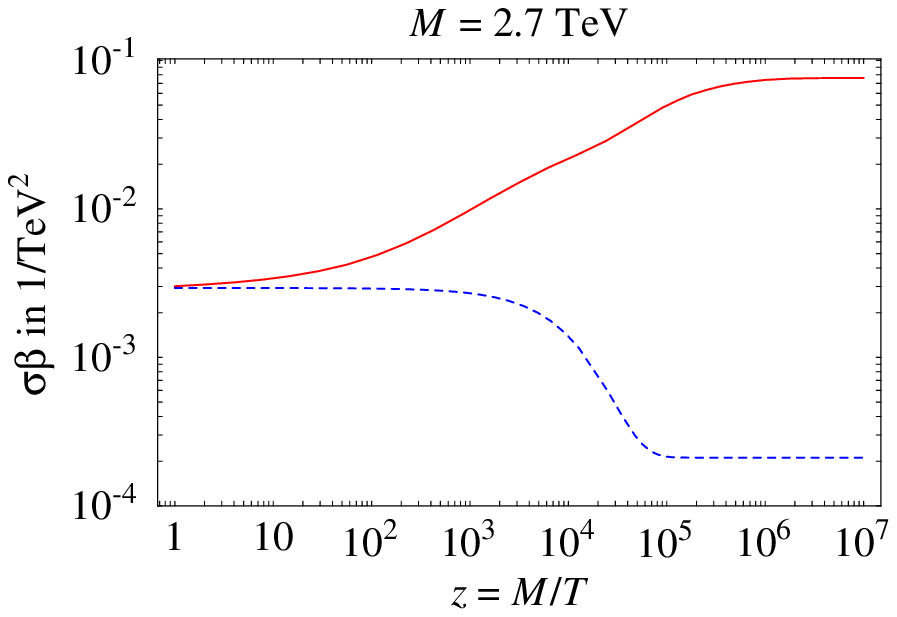}
\includegraphics[width=0.45\textwidth]{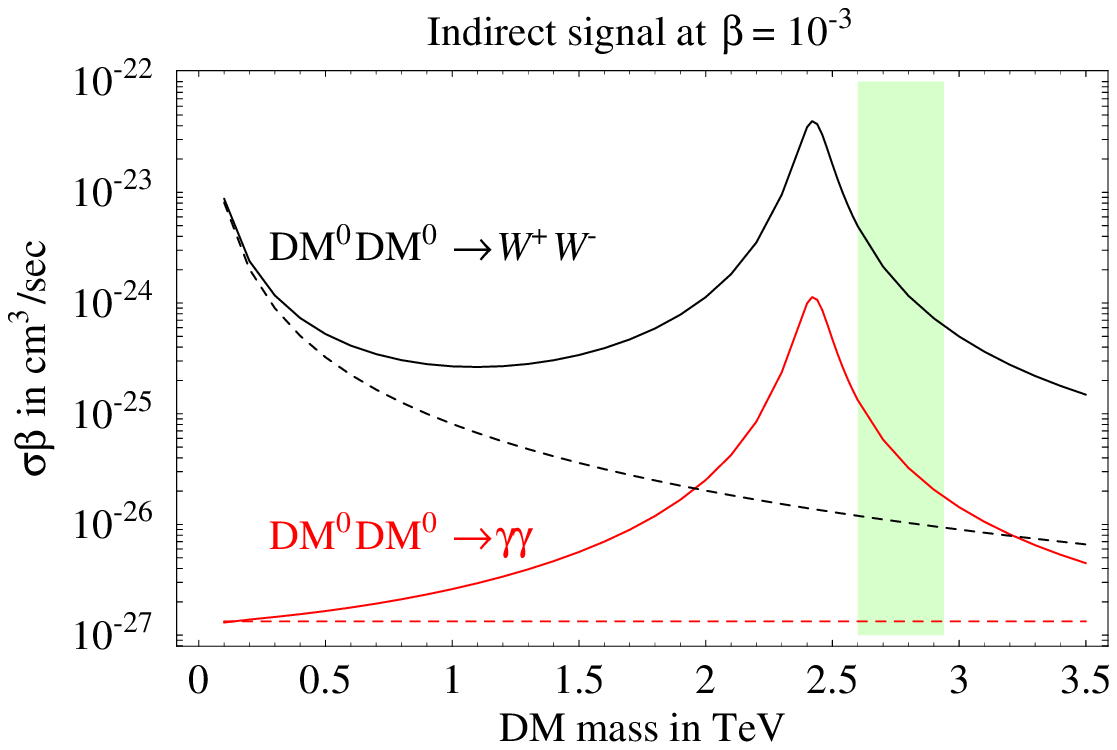}
\includegraphics[width=0.45\textwidth]{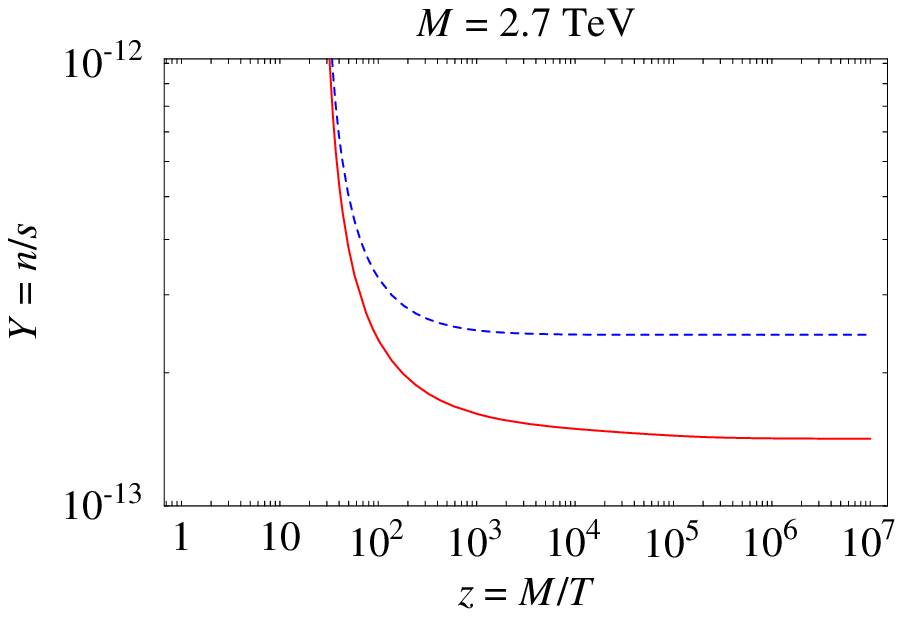}
\caption[X]{\label{fig:F30}\em {\bf The fermion triplet with zero hypercharge (`wino')}.
Fig.\fig{F30}a (upper left): our result for its cosmological freeze-out DM abundance as function of the DM mass.
Fig.\fig{F30}c (upper right) show an example (for the indicated mass $M$) 
of the temperature dependence of
the {\rm DM~DM} annihilation cross section,
and fig.\fig{F30}d (lower right) shows the resulting cosmological evolution of the DM abundance.
Fig.\fig{F30}b (lower left) shows our result for {\rm DM$^0$~DM$^0$} annihilation cross section
relevant for indirect DM detection, as discussed in section~\ref{indirect}.
In each case, the continuous line is our full result, while the dashed line is the result obtained
without including non-perturbative effects.}
\end{center}
\end{figure}

\subsubsection*{\rm\em The fermion triplet with $Y=0$ (`wino').}
It is not automatically stable: one needs to impose a suitable symmetry.
Since $Y=0$ the lightest component of this multiplet
is neutral under the $\gamma$ and under the $Z$, and therefore it 
satisfies bounds from direct DM searches~\cite{Xenon,MDM}.
This multiplet appears in supersymmetric models as $\SU(2)_L$ gauginos,
and behaves as MDM in limiting cases where it is much lighter than other sparticles.

We include $p$-wave DM~DM annihilations, but we only
include non perturbative corrections to $s$-wave annihilations, that are
splitted into four sectors
with $L=0$, total charge $Q=\{0,1\}$ and total spin $S=\{0,1\}$.
All sectors involve one component, except the sector with $Q=0$ and $S=0$,
that involves two components $\{{\rm DM}^-{\rm DM}^+, {\rm DM}^0{\rm DM}^0\}$.
In this case and in the following we indicate the basis we employ by writing the
charges of the DM components around the $\Gamma$ and $V$ matrices,
and by indicating $Q$ and $S$ as pedices and apices on $\Gamma$ and $V$.
Using eq.~(\ref{sys:GammaV}), for the fermion triplet we get:
\beq \Gamma_{Q=0}^{S=0} = \frac{\pi \alpha_2^2}{9M^2}\bordermatrix{&+&0\cr
-&3 & \sqrt{2}\cr0&\sqrt{2} & 2} ,\qquad
V_{Q=0}^{S=0}=\bordermatrix{&+&0\cr -&2\Delta-A&-\sqrt{2}B \cr 0& -\sqrt{2}B&0 },\eeq
\beq \Gamma_{Q=0}^{S=1}= \frac{25\pi\alpha_2^2}{36M^2}\qquad
V_{Q=0}^{S=1} = 2\Delta-A\eeq
\beq
\Gamma_{Q=1}^{S=0}= \frac{\pi \alpha_2^2}{9M^2},\qquad
\Gamma_{Q=1}^{S=1}= \frac{25\pi\alpha_2^2}{36M^2},\qquad
V_{Q=1}^{S=0,1}=\Delta-B   \label{eq:maledettosegno}\eeq
\beq\Gamma_{Q=2}^{S=0}=\frac{\pi \alpha_2^2}{9M^2},\qquad
V_{Q=2}^{S=0}=2\Delta+A .\eeq
where $\Delta=166\MeV$,
$A =   \alpha_{\rm em}/r + \alpha_2 c_{\rm W}^2 e^{-M_Zr}/r$ and 
$B=\alpha_2e^{-M_Wr}/r$.
These results are equivalent to those of~\cite{HisanoCosmo}.

We numerically solve the Schr\"{o}dinger equations
use the finite-temperature values for $\Delta$ and vector masses
(e.g.\ $\Delta$ vanishes when $\SU(2)_L$-invariance is restored, etc.).
Fig.\fig{F30}a shows our result for its freeze-out cosmological abundance: 
as previously noticed in~\cite{HisanoCosmo} non-perturbative corrections are relevant,
and significantly increase the multi-TeV
value of $M$ that reproduces the measured cosmological abundance.
In the supersymmetric case, a multi-TeV wino lightest supersymmetric particle
implies a fine-tuning in the Higgs mass above the $10^3$ level,
so that this scenario seems not motivated by the Higgs mass hierarchy problem.

A bound state first appears for $M > M_* \approx 2.5\TeV$,
in the DM DM system with total charge $Q=0$ and total spin $S=0$.
Fig.\fig{F30}c and d show more details of the computation for $M=2.7\TeV$. 
Since $M$ is just above $M_*$,
the bound state is loosely bound, $E_B \approx -67 \MeV$,
and gives rise 
to a significant non-perturbative
enhancement $R$ of the annihilation cross section, apparent in fig.\fig{F30}c as a dip
at $M=M_*$.
The result is reliable, because temperatures $T\circa{<}|E_B|$ 
(at which a dedicated treatment of bound states would be necessary)
do not significantly affect the final cosmological abundance, 
as indicated by  fig.\fig{F30}d and by the fact that the dip is not the main effect.


\subsubsection*{\rm\em The scalar triplet with $Y=0$.}
This MDM candidate is not automatically stable: one needs to impose a suitable symmetry.
Having $Y=0$, this candidate is compatible with bounds from direct DM searches.
Any SU(2)$_L$ multiplet of scalars
can have a quartic interaction  with the Higgs,
$-\lambda_H({\cal X}^\dagger T^a {\cal X}) (H^\dagger \tau^a H)$,
that generates a tree-level mass splitting within the multiplet.
Since this mass splitting is suppressed by the DM mass $M$~\cite{MDM}
it is not unbelievable that scalars behave as MDM because
$\lambda_H$ is small enough,
$\lambda_H \ll 0.05$, 
that the $\Delta=166\MeV$ mass splitting induced by gauge couplings is dominant.
This is particularly plausible in the case of scalars with $Y=0$,
because $\lambda_H$ is not generated by one-loop
RGE corrections. Indeed RGE corrections that induce a $\lambda_H$
proportional to $g_2^2$ happen to vanish, 
because the Higgs $H$ is a doublet, 
and the doublet representation of $\SU(2)_L$ has  vanishing symmetric tensor:
${\rm Tr}\, T^a_H\{T^b_H,T^c_H\}=0$.

To include non-perturbative corrections to
DM~DM annihilations we split them into three sectors,
with $L=0$, $S=0$ and total charge $Q=\{0,1,2\}$.
The $V$ and $\Gamma$ matrices are
\beq \Gamma_{Q=0}^{S=0} = \frac{2\pi \alpha_2^2}{9M^2}\bordermatrix{&+&0\cr
-&3 & \sqrt{2}\cr0&\sqrt{2} & 2} ,\qquad
V_{Q=0}^{S=0}=\bordermatrix{&+&0\cr -&2\Delta-A&-\sqrt{2}B \cr 0& -\sqrt{2}B&0 },\eeq
\beq \Gamma_{Q=1}^{S=0}= \frac{2\pi \alpha_2^2}{9M^2},\qquad
V_{Q=1}^{S=0}=\Delta +B ,\qquad
\Gamma_{Q=2}^{S=0}=\frac{2\pi \alpha_2^2}{9M^2},\qquad
V_{Q=2}^{S=0}=2\Delta+A .\eeq

\begin{figure}[t]
\begin{center}
\includegraphics[width=0.45\textwidth]{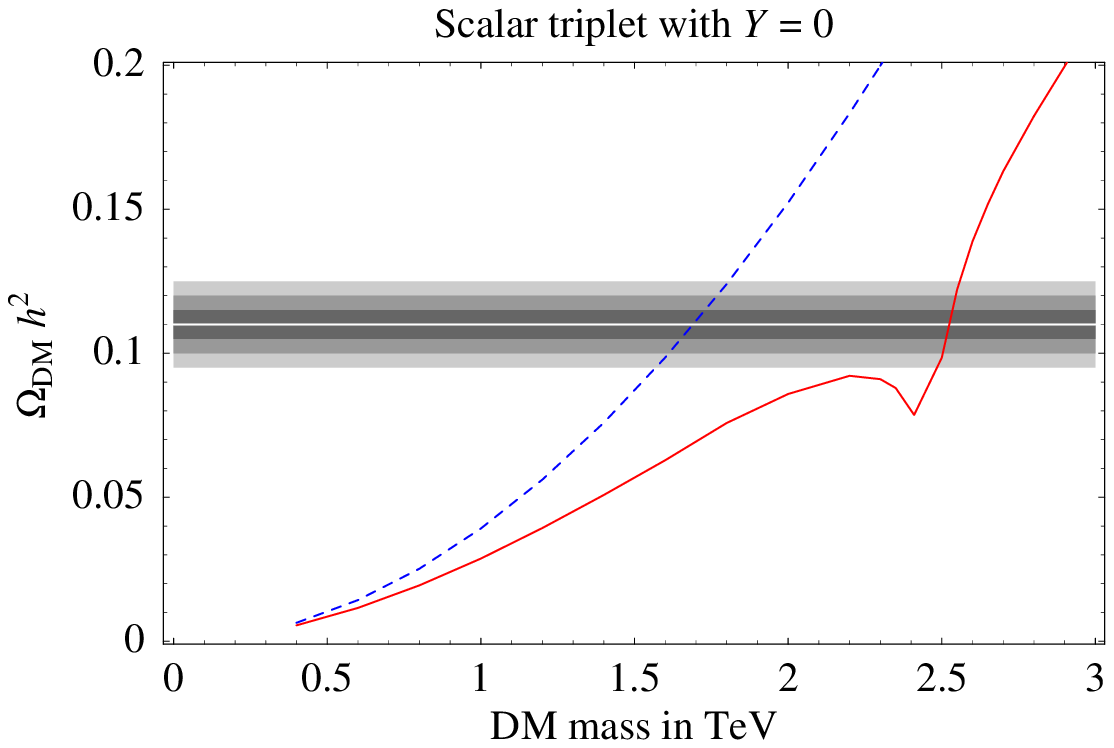}
\includegraphics[width=0.45\textwidth]{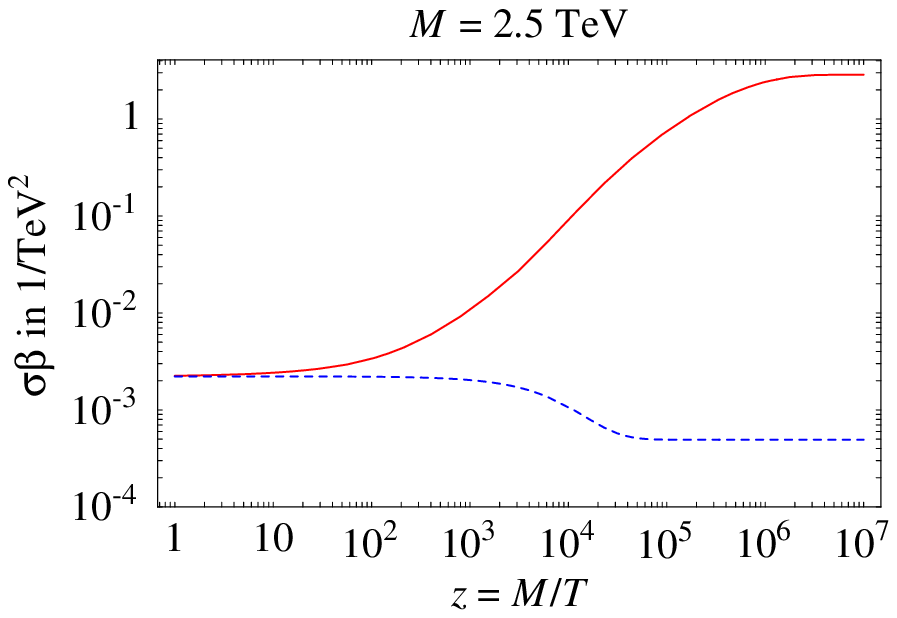}
\includegraphics[width=0.45\textwidth]{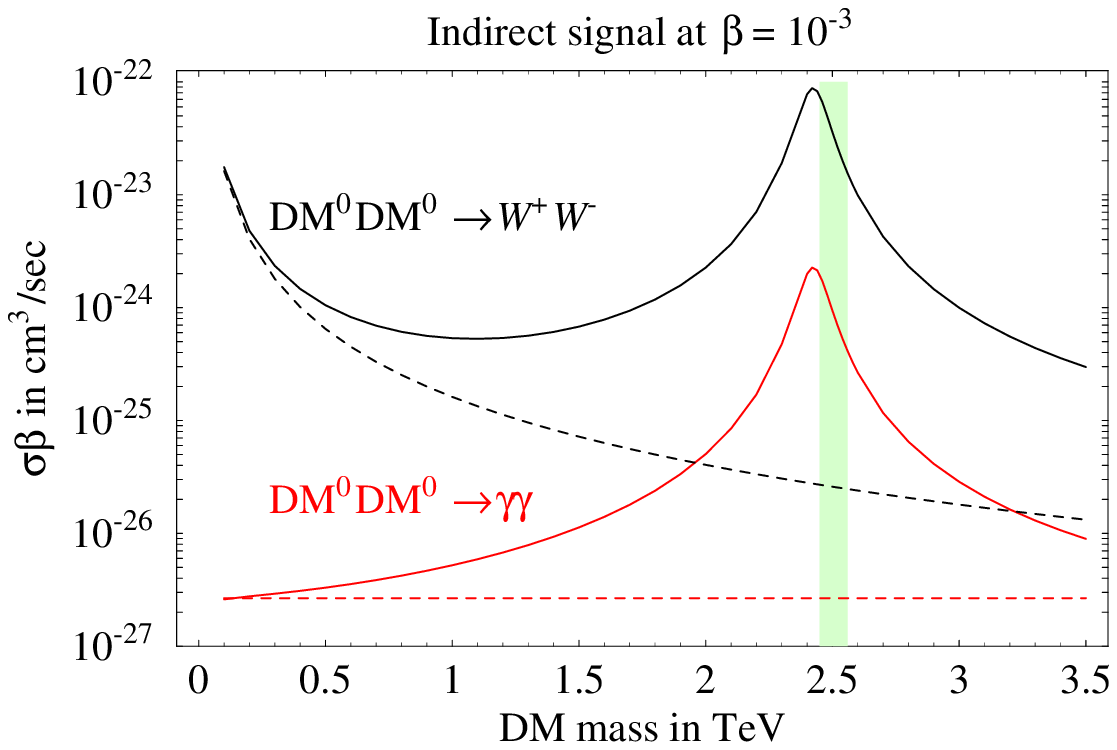}
\includegraphics[width=0.45\textwidth]{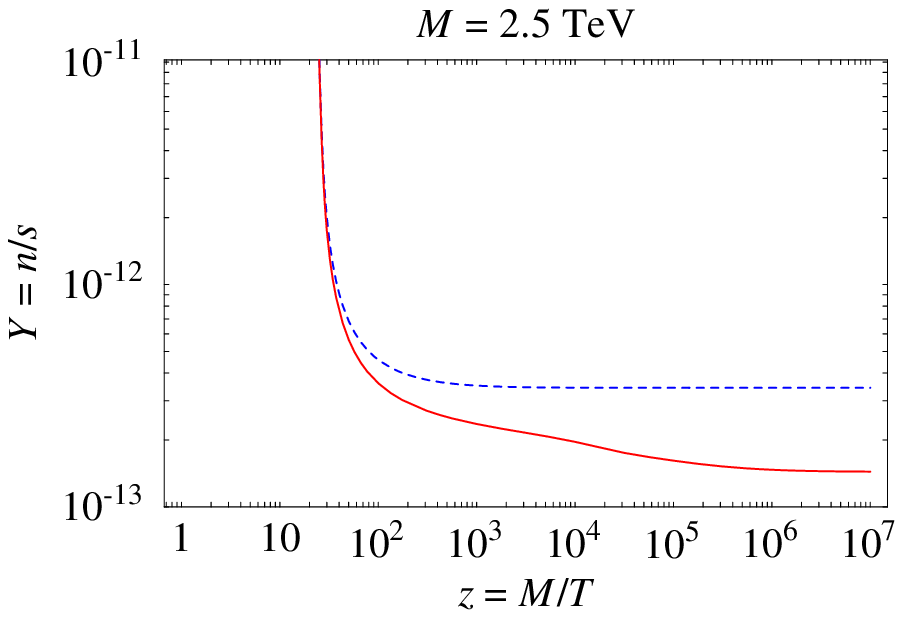}
\caption{\label{fig:S30}\em The scalar triplet with zero hypercharge.
Plots have the same meaning as in fig.\fig{F30}.}
\end{center}
\end{figure}

Fig.\fig{S30}a shows our result for its freeze-out cosmological abundance: 
non-perturbative corrections are again important.
Fig.\fig{S30}c and d show details of the computation for $M=2.5\TeV$,
a mass close to the critical mass $M_*\approx 2.5\TeV$ at which
a bound state appears in the DM DM system with total charge $Q=0$.

\medskip

We here do not study scalar and fermion triplets with $|Y|=1$;
the neutral component has $T_3 =Y\neq 0$ and consequently
couples to the $Z$, giving a too large direct detection rate.
A non-minimal mixing with a singlet is needed to avoid this problem.
One expects significant non-perturbative corrections, similarly
to the $Y=0$ case,
as  $\SU(2)_L$ gauge interactions are more significant than
U(1)$_Y$ the extra gauge interaction.

\begin{figure}[t]
\begin{center}
\includegraphics[width=0.45\textwidth]{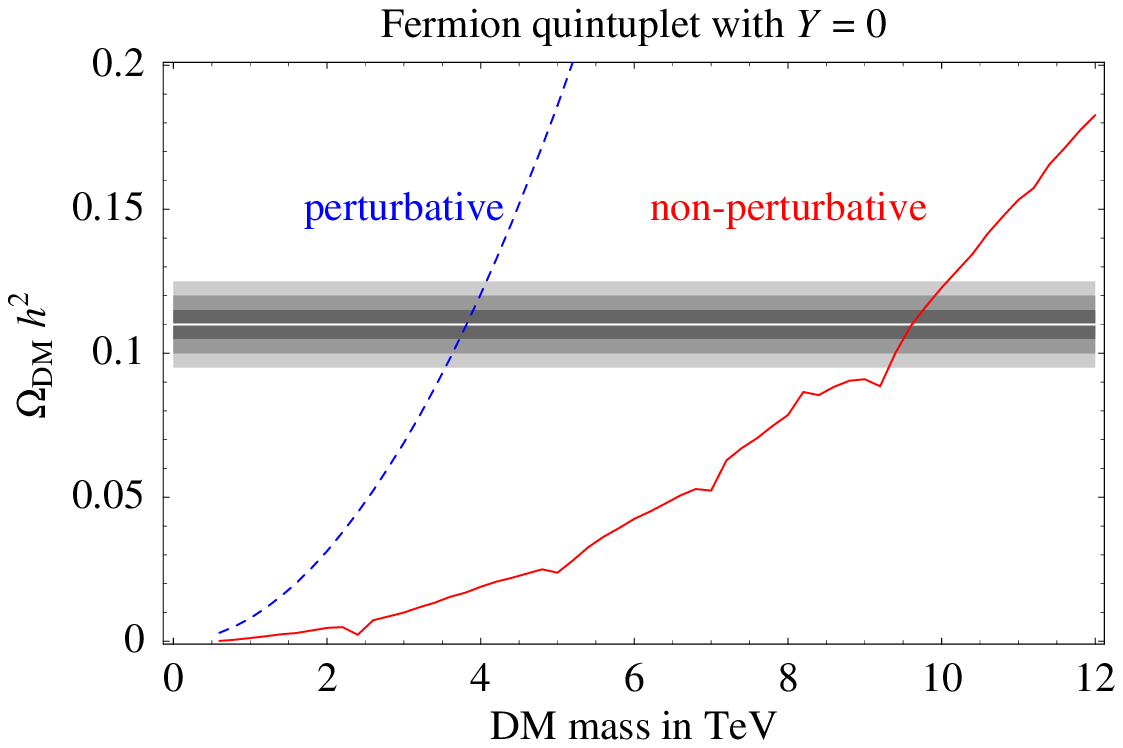}
\includegraphics[width=0.45\textwidth]{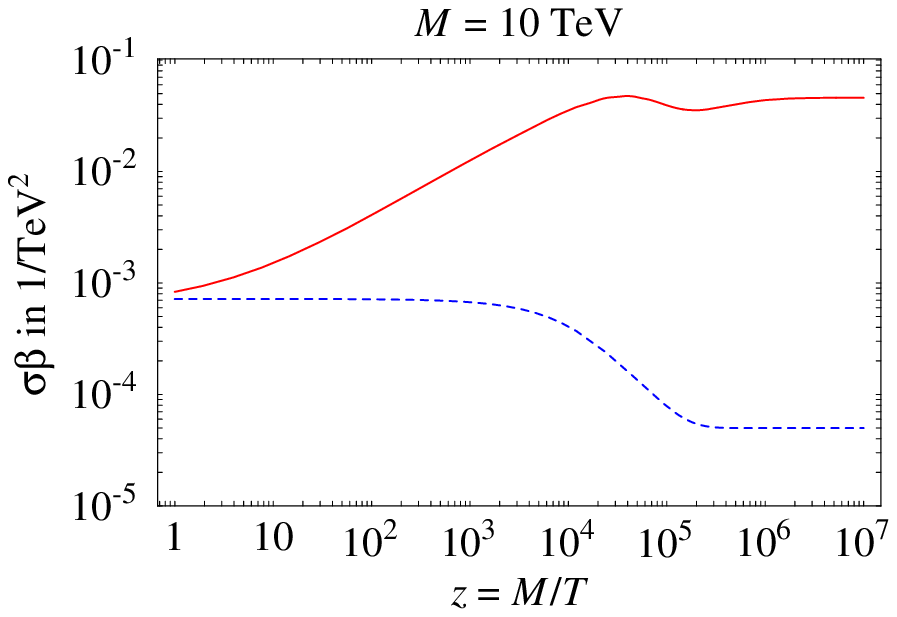}
\includegraphics[width=0.45\textwidth]{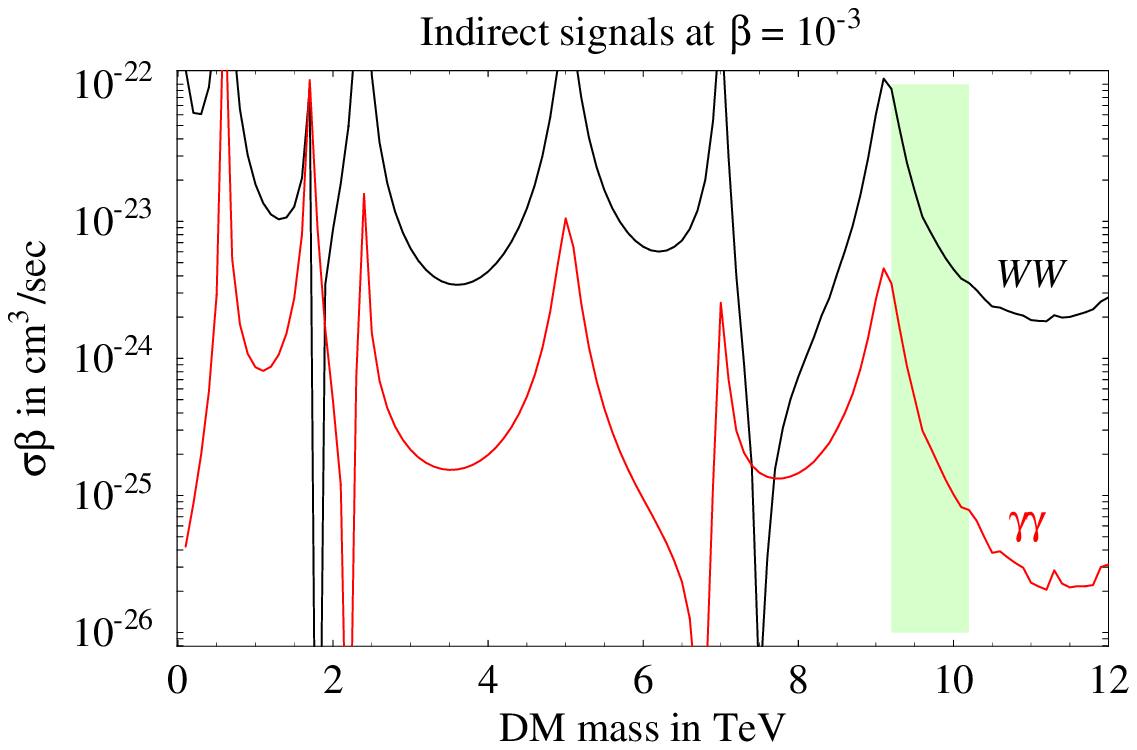}
\includegraphics[width=0.45\textwidth]{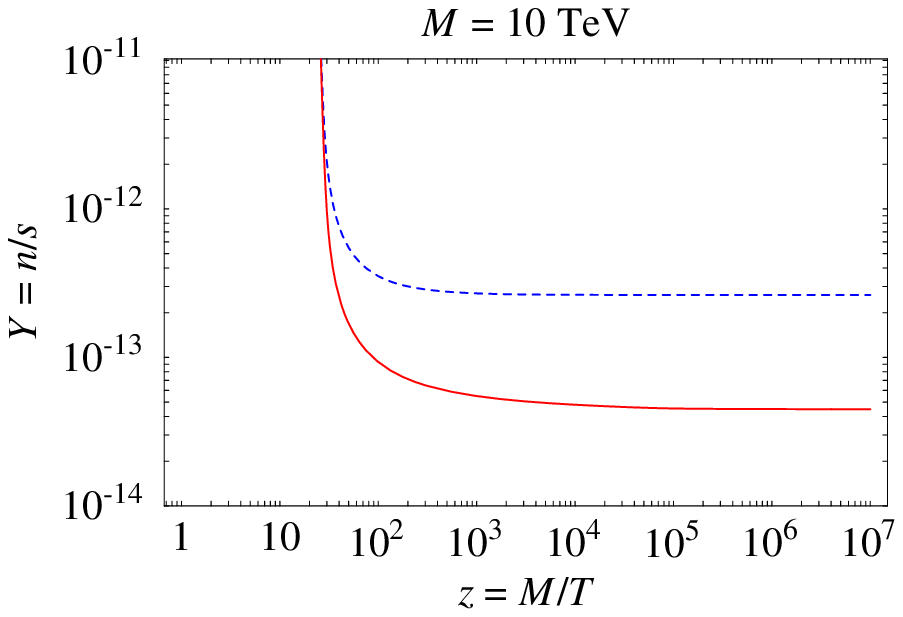}
\caption{\label{fig:F50}\em The fermion quintuplet with zero hypercharge.
Plots have the same meaning as in fig.\fig{F30}.}
\end{center}
\end{figure}

\subsubsection*{\rm\em The fermion quintuplet with $Y=0$.}
This is the smallest multiplet whose lightest component is 
neutral under the $\gamma$ and under the $Z$ and is automatically stable
on the cosmological timescale.
Indeed gauge couplings are the only renormalizable interactions
(compatible with gauge and Lorentz invariance)
 that this multiplet can have with other SM particles.
 Its phenomenology is univocally dictated by a single parameter: its Majorana mass $M$.
 
 To include non-perturbative corrections to
DM~DM annihilations we split these scatterings into four $L=0$ sectors,
with  total spin $S=\{0,1\}$ and total charge $Q=\{0,1,2\}$.
The $Q=0$ and $S=0$ system is composed by 3 states:
\beq
V_{Q=0}^{S=0} = \bordermatrix{&--&-&0\cr
++ &8\Delta -4A &- 2B &0\cr
+&-2B& 2\Delta - A&
-3\sqrt{2} B\cr
0&0&-3\sqrt{2} B & 0},\qquad
 \Gamma^{S=0}_{Q=0}=\frac{3\pi\alpha_2^2}{25M^2}\bordermatrix{&++&+&0\cr
-- & 12 &6 &2\sqrt{2}\cr
- &6 &9&5\sqrt{2}\cr
0 &2\sqrt{2}&5\sqrt{2}&6},
\eeq
Due to Fermi statistics,
the $Q=0$, $S=1$ system has no $\DM^0\DM^0$ state:
\beq
 V_{Q=0}^{S=1} = \bordermatrix{&--&-\cr
++ &8\Delta -4A &- 2B \cr
+&-2B& 2\Delta - A},\qquad
\Gamma^{S=1}_{Q=0}=\frac{\pi\alpha_2^2}{4M^2}\bordermatrix{&++&+\cr
-- & 4  &2\cr
- &2 & 1},\eeq
The systems with $Q=1$ and $S=\{0,1\}$ have two states each:
\beq V_{Q=1}^{S=0,1} = 
 \bordermatrix{& ++ & +\cr
- & 5\Delta  - 2 A & - \sqrt{6} B\cr
0 & -\sqrt{6} B&\Delta - 3B},
\eeq
\beq\Gamma^{S=0}_{Q=1}=
\frac{3\pi\alpha_2^2}{25 M^2}\bordermatrix{&++&+\cr
- & 6 &\sqrt{6}\cr
0 &\sqrt{6}&1\cr},\qquad
\Gamma^{S=1}_{Q=1}=
\frac{\pi\alpha_2^2}{4 M^2}\bordermatrix{&++&+\cr
- & 2 &\sqrt{6}\cr
0 &\sqrt{6}&3\cr}.
\eeq
Finally, in the systems with $Q=2$ the tree-level
annihilation rate $\Gamma$ is non vanishing only for $S=0$:
\beq V_{Q=2}^{S=0} = 
\bordermatrix{& ++ & +\cr
0 & 4\Delta &-2\sqrt{3}B \cr
+ &-2\sqrt{3}B&2\Delta +A},\qquad
\Gamma^{S=0}_{Q=2}=
\frac{3\pi\alpha_2^2}{25M^2}\bordermatrix{&++&+\cr
0 & 4 &-\sqrt{12}\cr
+ &-\sqrt{12}&3\cr}.\eeq
 Fig.\fig{F50}a shows our result for its freeze-out cosmological abundance: 
the value of $M$ that reproduces the measured DM abundance 
was $M\approx 4.4\TeV$. It slightly decreases down to $M\approx 3.8\TeV$ after including
$p$-wave and RGE corrections (perturbative result in fig.\fig{F50}a),
and increases up to almost $M\approx 10\TeV$ after including the non-perturbative Sommerfeld  corrections.
Fig.\fig{F50}d shows that Sommerfeld corrections are
mostly relevant when $T \gg \Delta M$, such that 
bound states are not expected to play a relevant r\^ole.

%


\subsubsection*{\rm\em The scalar quintuplet with $Y=0$.}
 As in the scalar triplet case, is
plausible to assume that the quartic coupling with the Higgs is negligibly small,
because it is not  generated by RGE effects.
The $V$ and $\Gamma$ matrices are related to those relevant for the 
fermion 5-plet as described in section~\ref{somm}.
Non-perturbative corrections increase
the value of the mass $M$ that reproduces the cosmological abundance from
$M\approx 5.0\TeV$~\cite{MDM} to $M\approx 9.4\TeV$.

\subsubsection*{\rm\em The scalar eptaplet with $Y=0$.}
 This is the smallest scalar multiplet that is automatically stable
(its cubic scalar coupling identically vanishes)
 and contains a MDM candidate compatible with all bounds.
Again, it is
plausible to assume that the quartic coupling with the Higgs is negligibly small.
Although we do not show dedicated plots nor the $V$ and $\Gamma$ matrices,
we computed non-perturbative corrections finding that they increase
the value of the mass $M$ that reproduces the cosmological abundance from
$M\approx 8.5\TeV$~\cite{MDM} to about $M\approx 25\TeV$.

\begin{figure}[t]
\begin{center}
\includegraphics[width=0.45\textwidth]{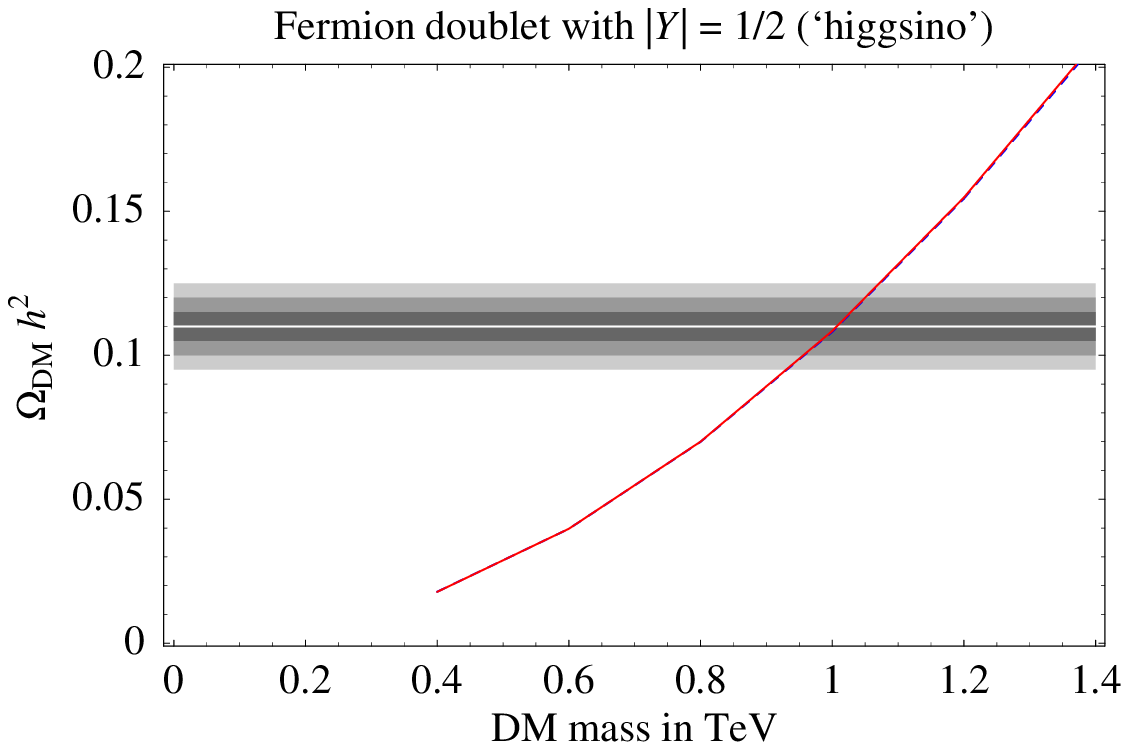}
\includegraphics[width=0.45\textwidth]{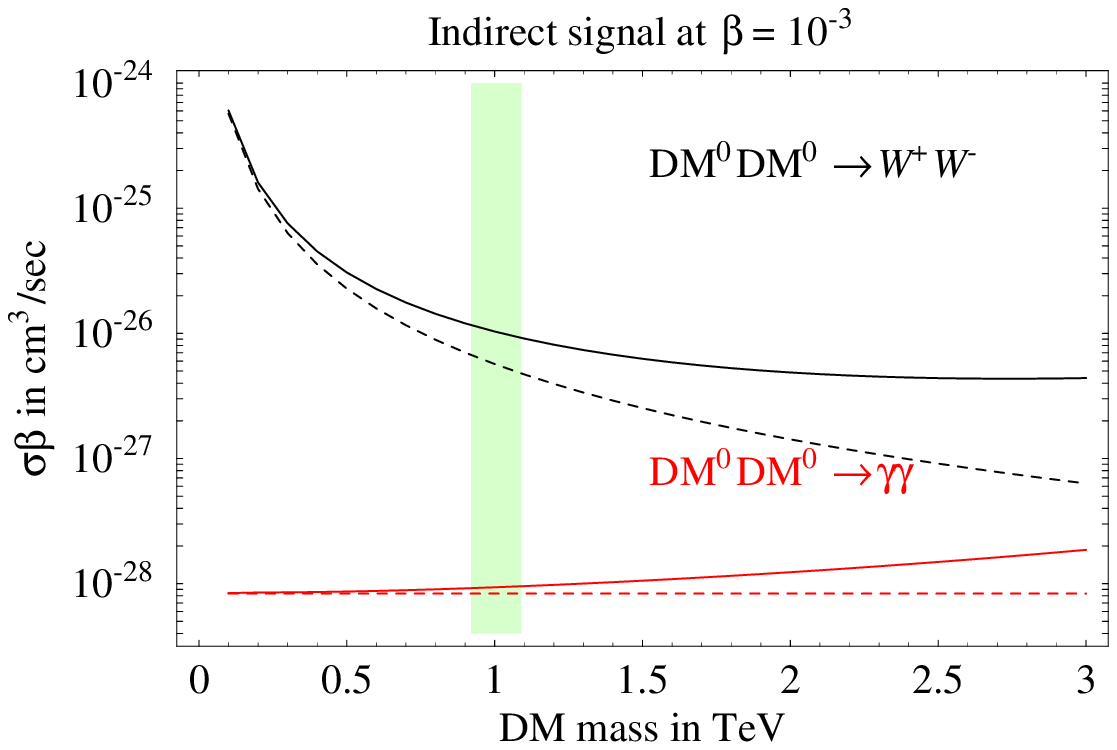}
\caption{\label{fig:F2}\em Cosmological freeze-out abundance of 
the fermion doublet with $Y=1/2$ (`Higgsino').
Plots have the same meaning as in fig.\fig{F30}, except that since
non-perturbative corrections negligibly affect the cosmological abundance
we do not show details of the computations.
}
\end{center}
\end{figure}

\subsubsection*{\rm\em The fermion doublet with $Y=1/2$ (`Higgsino').}
It is not automatically stable, and, having $Y\neq 0$, is excluded by direct DM searches.
Nevertheless, we consider it because $n=2$ is the smallest multiplet, 
and because it is present in supersymmetric models.
The first problem can be solved by imposing a suitable symmetry, and incompatibility with direct detection experiments can be avoided by assuming a mass mixing with 
a neutral Majorana singlet (automatically present in supersymmetric models and known as `bino').
We assume that this mass mixing is small enough not to affect the observables we compute.
An almost-pure Higgsino LSP is realized in corners of the MSSM parameter space:
it behaves as Minimal Dark Matter in the sense that its properties
are not dictated by supersymmetry but only by gauge invariance.
In this limit, the $\mu$-term is the only relevant supersymmetric parameter, and coincides with our
DM mass parameter $M$.

To include non-perturbative corrections to
DM~$\overline{\rm DM}$ annihilations we split these scatterings into four $L=0$ sectors,
with total charge $Q=\{0,1\}$ and total spin $S=\{0,1\}$.
The $\Gamma$ and $V$ matrices that describe each sector are 
$$\Gamma_{Q=0}^{S=0} = \frac{\pi \alpha_2^2}{64 M^2}  \bordermatrix{&+&\bar 0\cr
-&3+2t^2+t^4 & 3-2t^2+t^4\cr
0&3-2t^2+t^4 & 3+2t^2 + t^4},\qquad
\Gamma_{Q=0}^{S=1} = \frac{\pi \alpha_2^2}{128 M^2}  \bordermatrix{&+&\bar 0\cr
-&41t^4+25 & 41t^4-25\cr
0&41 t^4-25&41 t^4+25},$$
\beq V_{Q=0}^{S=0,1}= 
\bordermatrix{&+&\bar 0\cr
-&-\alpha_{\rm em}/r -(2c^2-1)^2 e^{-M_Z r}/4rc^2&-\alpha_2 e^{-M_Wr}/2r\cr
0&-\alpha_2 e^{-M_Wr}/2r&-e^{-M_Z r}\alpha_2/4rc^2}\eeq
\beq \Gamma_{Q=1}^{S=0}= \frac{\pi \alpha_2^2 t^2}{16M^2},\qquad
\Gamma_{Q=1}^{S=1} = \frac{25\pi \alpha_2^2}{64 M^2},\qquad
V_{Q=1}^{S=0,1}=+ \frac{\alpha_2 e^{-M_Z r}}{r }\frac{2c^2-1}{4c^2}
\eeq
where $t = \tan\theta_{\rm W}$, $c=\cos\theta_{\rm W}$ and
$\Delta=341\MeV$ is the mass splitting among charged and neutral components
generated by loop effects.
Fig.\fig{F2}a shows our result for its cosmological abundance:
in agreement with previous studies (e.g.~\cite{MDM})
we see that the observed DM abundance is reproduced for
$M=1\TeV$. Non perturbative corrections are negligible.
This also holds for the scalar doublet with $|Y|=1/2$, so we do not study it.

The Higgs potential in the Minimal Supersymmetric Standard Model
depends on the $\mu$ parameter: 
a $|\mu| = 1\TeV$ gives a contribution to the squared $Z$-mass which is 
$2\mu^2/M_Z^2=240$ times too large:
therefore the  Higgsino as Minimal Dark Matter
can only be realized at the price of a sizable fine tuning.

\section{Indirect detection of Minimal Dark Matter}\label{indirect}
We now study the usual ``indirect DM signals'', generated by DM$^0$ DM$^0$
annihilations into SM particles. 
In the MDM case one only has tree-level annihilations into $W^+W^-$,
while the most interesting  final states, $\gamma\gamma$ and
$\gamma Z$ (both detectable as $\gamma$ with energy practically equal to $E_\gamma =M$),
arise at loop level.
Assuming $Y=0$ the cross sections (averaged over DM polarizations) are, in the fermion DM case
\beq \label{eq:20a}
\sigma(\DM^0\,\DM^0\to W^+ W^-) \beta = (n^2-1)^2\frac{\pi \alpha_2^2}{32M^2},\qquad
\sigma(\DM^0\,\DM^0\to \gamma\gamma) \beta = (n^2-1)^2\frac{\pi \alpha^2_{\rm em}\alpha_2^2}{16 M^2_W}.
\eeq
and 2 times higher in the scalar DM case.
These cross-sections can be significantly affected by non-perturbative Sommerfeld corrections~\cite{Hisano}, because in our galaxy DM has a non-relativistic velocity $\beta\approx 10^{-3}$.
The computation employs the same tools already described in the computation
of DM DM annihilations relevant for cosmology.
Non-perturbative effects can enhance
the cross section  by orders of magnitude for specific values of $M$:
the values at which the potential develops a new bound state with $Q,L,S=0$.
In the triplet and quintuplet cases this enhancement is sizable, since
the DM  mass suggested by cosmology is  close to one of such critical values. 
We show our results for the fermion triplet in fig.\fig{F30}b,
for the scalar triplet in fig.\fig{S30}b,
for the fermion quintuplet in fig.\fig{F50}b, and for the fermion doublet in fig.\fig{F2}b.
The vertical bands are the $3\sigma$ range of $M$ suggested by cosmology.
We have not plotted annihilation cross sections into $\gamma Z$ and $ZZ$ since
all MDM candidates with $Y=0$ predict
\beq \sigma_{\gamma Z} = 2\sigma_{\gamma\gamma}/\tan^2\theta_{\rm W}=
6.5 \sigma_{\gamma\gamma}
,\qquad
\sigma_{ZZ} = \sigma_{\gamma\gamma}/\tan^4\theta_{\rm W} = 10.8\sigma_{\gamma\gamma}.\eeq

\begin{figure}[t]
\begin{center}
\includegraphics[width=0.5\textwidth]{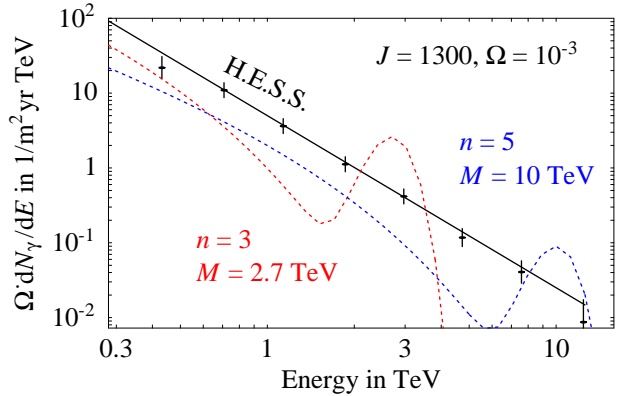}
\caption{\em\label{fig:gamma} Spectrum of $\gamma$ from
$\DM^0 \DM^0$ annihilations around the galactic center, in a region of size $\Omega=10^{-3}$,
as seen by a detector with $\sigma_E/E=0.15$.
The total rate is computed  assuming the NFW profile, $J=1300$,
and for fermion MDM with $n=3$ and $M=2.7\TeV$, and for
$n=5$ and $M=10\TeV$.
The continuous line is the H.E.S.S.\ result~\cite{HESS}.
}
\label{default}
\end{center}
\end{figure}

These results allow to compute the energy spectra of annihilation products, such as
photons,  antiprotons, antideutrons and positrons.
Note however that, precisely for the proximity to one of the critical values in mass, the cross sections depend strongly on the DM mass $M$ and therefore the overall fluxes cannot be accurately predicted. 
Moreover, such total rates are affected by  sizable astrophysical uncertainties.
For example, the number of detected photons with energy $E=M$  is~\cite{Ullio}
\beq 
N_\gamma (\hbox{at }E=M) \approx \frac{30}{{\rm m}^2\,{\rm yr}\,{\rm sr}} J\,\Omega
  \left( \frac{\TeV}{M} \right)^2 \frac{2\sigma_{\gamma\gamma} \beta+\sigma_{\gamma Z}\beta}{10^{-24}\,{\rm cm}^3/{\rm s}} .
\eeq
where $\Omega$ is the angular size of the observed region
and the quantity $J$ ranges between a few and $10^5$,
depending on the unknown DM density profile in our galaxy.
Keeping these limitations in mind, fig.\fig{gamma} shows the predicted energy spectrum of detected photons from MDM annihilations. We have assumed realistic detector parameters
($\Omega = 10^{-3}$ and an energy resolution of $15\%$).
The contribution from annihilations into $W^+W^-$
is included using the spectral functions computed in~\cite{Ullio,Hisano}.
The total rate has an uncertainty of about two orders of magnitude and is here
 fixed assuming the indicated value of $M$
and the Navarro-Frenk-White DM density profile, $J =1300$~\cite{NFW,DMreview},
which makes this signal at the level of the present sensitivity.
Indeed the continuous line shows the spectrum of diffuse $\gamma$,
${dN_\gamma}/{dE} \approx  5~10^3({\TeV}/{E})^{2.3} /{{\rm m}^2{\rm yr}~\TeV~{\rm sr}} $,
as observed (in a region of angular size $\Omega \approx 10^{-3}$ around the galactic center) by H.E.S.S~\cite{HESS} up to energies of about 10 TeV,
where the number of events drops below one.
Its spectral index suggests that the observed photons are generated
by astrophysical mechanisms, rather than by DM DM annihilations.
In view of the predicted value of $\sigma_{\gamma\gamma}/\sigma_{WW}$,
the best MDM signal is the peak at $E_\gamma = M$.
With the chosen parameters one gets
$N_\gamma = 0.3~(2.6)/{\rm m^2}\cdot{\rm yr}$ for the fermion 3-plet (5-plet).


\section{Minimal Dark Matter at Ultra High Energies}\label{Astro}
In this section we discuss the possibility that Minimal Dark Matter might be detected via the tracks left by its charged partners in experiments mainly devoted to Cosmic Ray (CR) detection, such as 
 IceCUBE~\cite{ICECUBE} and km3net~\cite{Antares}.
Indeed MDM behaves differently from other neutral particles:
\begin{itemize}
\item Neutrinos  with energy $E\circa{<}10^{15}\eV$  can
cross the whole Earth ~\cite{nutransport} and, impinging on the rock or ice below the detector, produce a charged partner (a muon, a tau) that cruises the instrumented volume. 

\item 
Neutral particles much heavier than neutrinos (e.g.\ the speculative neutralinos)
interact rarely enough that they can cross the Earth even at Ultra High Energies (UHE),
$E\gg 10^{15}\eV$~\cite{neutralinotransport}. 
If they do interact before the end of their journey, they do produce a charged partner
(e.g.\ the speculative chargino), that however usually decays almost immediately back to the neutral particle. For such reasons, it is generically difficult to detect them in this way.
\end{itemize}
MDM is both heavy and quasi-degenerate with a charged partner, leading to the following behavior.
To be quantitative, we focus on MDM multiplets  with $Y=0$.
They have a  stable neutral DM particle, $\DM^0$,
and  a charged $\DM^\pm$ component which is $166\MeV$ heavier.
The latter dominantly decays via $\DM^\pm \to \DM^0 \pi^\pm$
with a macroscopic life-time $\tau = 44\,{\rm cm}/(n^2-1)$~\cite{MDM}.
At Ultra High Energy Lorentz dilatation makes the $\DM^\pm$ life-time
long enough that a sizable fraction of the travel is done as $\DM^\pm$,
leading to detectable charged tracks.\footnote{Supersymmetric models with gravitino LSP may have a long-lived,  charged next-to-lightest particle that behaves in a way qualitatively similar to MDM~\cite{longlivedSuSy}.  
Furthermore, fermionic MDM with $n=3$ and $Y=0$ arises in supersymmetry in the limit of pure-wino lightest supersymmetric particle.}



This opens in principle an interesting and distinctive channel for the detection of DM.
In order to assess its feasibility in practice, we need to study the system of MDM from production to detection.
More precisely, first we study the issue of how a MDM system behaves when crossing the Earth at UHE, having assumed that the flux of UHE CR (observed up to $E\sim 10^{20}\GeV$) does contain some DM particles.
Mechanisms to produce such UHE DM are discussed later in section~\ref{UHEprod}. 
We finally discuss detection signatures in section~\ref{UHEdetection}.

\subsection{Propagation of UHE Minimal Dark Matter}
\label{UHEpropagation}

Consider a flux of $\DM^0$ particles that is crossing the Earth at UHE. Via Charged Current (CC) interactions with nucleons of Earth's matter, $\DM^\pm$ particles are produced. Being charged particles traveling in a medium, these loose a part of their energy and eventually decay back to $\DM^0$ particles. 
This chain of production and decay is analogous to the process that tau neutrinos undergo in matter (``$\nu_\tau$-regeneration"). In the $n=5$ case $\DM^{\pm\pm}$ also play a role, but we will write
explicit equations for the $n=3$ (wino-like) case.
Such a system is described by a pair of coupled integro-differential equations for the evolution with the position $\ell$ of fluxes $n_0=dN_0/dE$ and $n_\pm=d(N_+ + N_-)/dE$ of $\DM^0$ and ($\DM^+$ + $\DM^-$). They read:\footnote{Similar equations have been of course used in the analogous problem for neutrinos~\cite{nutransport} and for high energy neutralinos~\cite{neutralinotransport}.}
\begin{eqnsystem}{sys:evoDM}
 \frac{dn_0(\ell,E)}{d\ell} &=&+\frac{M}{E} \frac{n_\pm(E)}{\tau} + N_N\bigg[- n_0(E)\sigma_{\rm CC}(E)+\frac{1}{2}\int_E^\infty  dE' n_\pm(E')  \frac{d\sigma_{\rm CC}(E',E)}{dE}\bigg],\\
 \frac{dn_\pm (\ell,E)}{d\ell} &=& -\frac{M}{E} \frac{n_\pm(E)}{\tau}+N_N\bigg[
- \frac{1}{2}  n_\pm(E) \sigma_{\rm CC}(E) +   \nonumber
\int_{E}^\infty  dE'
 n_0(E')\frac{d\sigma_{\rm CC}(E',E)}{dE} \bigg]+ \\&&-
\frac{d}{dE}( n_\pm  \frac{dE_{\rm loss}}{d\ell}),
 \end{eqnsystem}
where $N_N(\ell)$ is the number density profile of nucleons  in the Earth~\cite{PREM},
and $E$ is the DM energy.
The first terms account for $\DM^\pm$ decays. 
The integro-differential terms within square brackets account for
$\DM^0\leftrightarrow\DM^\pm$ CC scatterings.
The last term accounts for the energy losses undergone by $\DM^\pm$,
due to electromagnetic and Neutral Current (NC) scatterings:
\beq
\frac{dE_{\rm loss}}{d\ell} = \frac{dE_{\rm loss,\, em}}{d\ell} +\frac{dE_{\rm loss,\, NC}}{d\ell} .
\eeq
We will later verify that the NC energy losses are subdominant and can be approximated as
continuous, like electromagnetic energy losses.

\medskip

The cross section for the CC partonic scatterings
$\DM^0~d \to \DM^-~u$, $\DM^0 \bar{d} \to \DM^+~\bar{u}$,
$\DM^0~ u \to \DM^+~d$ and $\DM^0 \bar{u} \to \DM^-~ d$
are all given by, for fermionic MDM:
\beq
\frac{d\hat\sigma_{\rm CC}}{d\hat{t}}=
\left( \frac{E}{(\hat s - M^2)}\frac{d\hat\sigma_{\rm CC}}{dE} = \right)
\frac{g_2^4(n^2-1)}{256\pi}
\frac{2M^4 + 2 \hat{s}^2 + 2\hat{s}\hat{t}+\hat{t}^2 - 4M^2 \hat{s}}{(\hat{s}-M^2)^2(\hat{t}-M_W^2)^2}
\eeq
where $\hat{s}$ and $\hat{t}$ are the usual partonic Mandelstam variables.
The total cross section for $\DM^0 N \to \DM^+ N',\DM^- N'$ is
\beq 
\sigma_{\rm CC}(s) = 
\int_0^1 dx ~q(x) \int_{-(\hat{s}-M^2)^2/\hat{s}}^0 d\hat{t} ~ \frac{d\hat\sigma}{d\hat t},
\eeq
where $q(x)$ is the parton distribution function~\cite{PDF}, summed over all quarks and anti-quarks.
We here approximated the neutron/proton fraction in Earth matter as unity.
The nucleon $N$ is at rest with mass $m_N$, and $\DM^0$ has energy $E_0$, so that
$s=M^2 + 2m_N E_0$ and $\hat{s} = M^2 + 2x m_N E_0$.
A similar expression holds for NC scatterings of $\DM^\pm$ particles $\DM^\pm q \to \DM^\pm q$ (whereas $\DM^0$ has no NC interactions due to $Y=0$):
\beq \frac{d\hat{\sigma}_{\rm NC}}{d\hat{t}}  =
\frac{g_2^4(g_{Lq}^2+g_{Rq}^2)}{16\pi}
\frac{2M^4 + 2 \hat{s}^2 + 2\hat{s}\hat{t}+\hat{t}^2 - 4M^2 \hat{s}}{(\hat{s}-M^2)^2(\hat{t}-M_Z^2)^2}\eeq
where $g_{Lq} = T_{3q} - Q_q s_{\rm W}^2$,
$g_{Rq} = -Q_q s_{\rm W}^2$
and $q$ denotes any quark or anti-quark.
The same cross sections, up to terms suppressed by powers of $M_{W,Z}/M$,
hold for scalar MDM.

\begin{figure}[t]
\begin{center}
\includegraphics[width=0.98\textwidth]{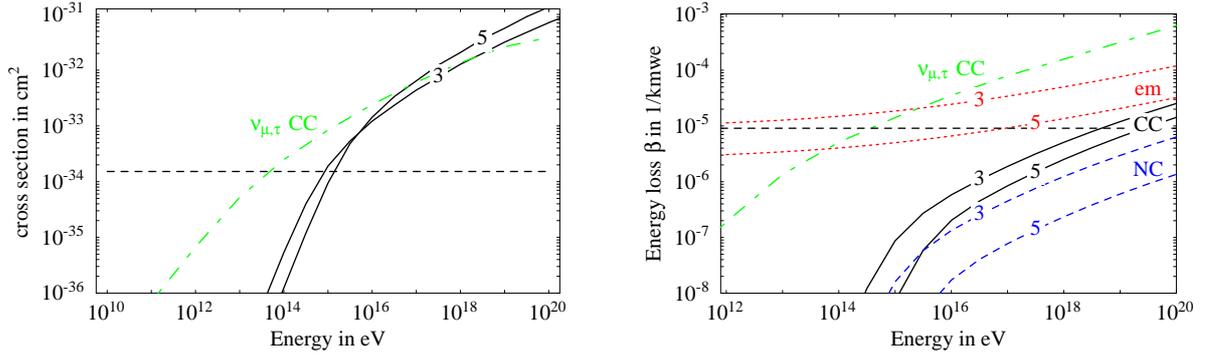}
\caption{\label{fig:Eloss}\em Left plot: CC cross section for fermionic MDM (black solid) with $n=3,5$ on nucleons. The same cross section for neutrinos (green dot-dashed) is reported for comparison.
Right plot: average energy loss parameters $\beta= - d\ln E/d\ell$ 
due to CC interactions (black solid lines), NC interactions (blue dashed), 
and electromagnetic interactions (red dotted). 
For comparison,
the green dot-dashed line show the CC energy loss of $\nu_{\mu,\tau}$.
The horizontal lines show the thickness of the Earth, crossed vertically.}
\end{center}
\end{figure}

\medskip

In fig.\fig{Eloss}a we plot the MDM total CC cross section. We can understand its main features as follows.
The total partonic cross section approximatively is
\beq \hat\sigma_{\rm CC}\simeq \frac{g_2^4(n^2-1)}{128\pi M_W^2}\cdot
\left\{\begin{array}{ll}
2 (xE/E_*)^2 & \hbox{at $xE\ll E_*$}\\
1 & \hbox{at $xE\gg E_*$}
\end{array}\right.\eeq
where $E_* \equiv M M_W/m_N \sim 10^{15}\eV$
(the transition between the two regimes happens when $t\sim - M_W^2$).
Therefore one has a constant $\hat\sigma_{\rm CC}\approx 10^{-33}\,{\rm cm}^2$ at $xE\gg E_*$.
Upon integration over the parton densities, the nucleonic $\sigma_{\rm CC}$ becomes a slowly growing function of $E$ because at higher $E$ partons with smaller $x\circa{>}E_*/E$ contribute,
and $q(x)$ grows as $x\to 0$.
This is analogous to what happens for neutrinos~\cite{review}.
%
Indeed, at high energies, the MDM CC cross section in fig.\fig{Eloss} approaches the dotted red line of the analogous result for $\nu_{\mu,\tau}$.

\medskip

For the electromagnetic energy losses of $\DM^\pm$ one has the standard parameterization~\cite{PDG}
\begin{equation}
-\frac{dE_{\rm loss,\, em}}{d\ell} = \alpha +\beta_{\rm em} E
\end{equation}
where $\alpha \approx 0.24\TeV/{\rm kmwe}$ approximates the energy loss due to
ionization effects~\cite{PDG} and the $\beta_{\rm em}$ term accounts for the radiative losses: it receives contributions from bremsstrahlung, $e^+e^-$ pair production and photonuclear scattering (i.e. inelastic electromagnetic collisions on nuclei). The latter two effects have comparable importance, while bremsstrahlung is subdominant for a very heavy particle such as MDM~\cite{Reno}. We follow the approach of~\cite{Reno} and adopt a parameterization for $\beta_{\rm em}$ with a mild dependance on the energy:
\begin{equation}
\beta_{\rm em} \simeq 2.5\ 10^{-5} \frac{1}{{\rm kmwe}}\frac{\TeV}{M}\left[1+\left( \frac{E}{10^{15}\eV} \right)^{0.215} \right].
\end{equation}


\bigskip

Before proceeding to the numerical solution of the equations in (\ref{sys:evoDM}), we can gain  understanding on the expected results by simplifying the treatment of the CC energy losses. In analogy with the standard parameterization, we can define an average energy loss suffered by $\DM^0$ particles due to CC interactions as
\beq
-\beta_{\rm CC}(E)= \frac{1}{E} \frac{dE_{\rm loss,\, CC}}{d\ell} = N_N
 \int_0^1 dx ~q(x) \int_{-(\hat{s}-M^2)^2/\hat{s}}^0 d\hat{t} ~\frac{\hat{t}}{\hat{s}-M^2}~ \frac{d\hat\sigma_{\rm CC}}{d\hat t} .\eeq
 A completely analogous expression holds for $\beta_{\rm NC} = - 1/E\ dE_{\rm loss,\, NC}/d\ell$. 
Both are plotted in fig.\fig{Eloss}b, in units of $1/{\rm kmwe}= 10^{-5}{\rm cm}^2/{\rm gram}$. The horizontal line indicates the value above which a particle looses a significant amount of energy
while vertically crossing the Earth (its thickness being $1.1~10^5\,{\rm kmwe}$).
Energy losses can be understood analogously to the total cross section, 
and are dominated by partons with $xE\sim E_*$.
Parton distributions have been measured at $x\circa{>} 10^{-4}$ and 
theoretical extrapolations to smaller $x$ can have ${\cal O}(1)$ uncertainties,
that affect our results.

The fraction of energy lost in one scattering is small,
$\sim \beta/N_N\sigma \circa{<}10^{-2}$
at the UHE energies that will be relevant for us, so
that all energy losses can be approximated as continuous:\footnote{The fraction of energy lost in one $\DM^\pm \to \DM^0$ decay
is $\Delta M/M \sim 10^{-4}$ and can be neglected.}
\beq -\frac{dE_0}{d\ell} = \beta_{\rm CC} E_0,\qquad
-\frac{dE_\pm}{d\ell} = \alpha + (\frac{\beta_{\rm CC}}{2}+\beta_{\rm em}+\beta_{\rm NC}) E_\pm\eeq
where $E_0$ ($E_\pm$) is the energy of $\DM^0$ ($\DM^\pm$).
Therefore, we do not need to study the evolution of the $\DM$ energy spectrum 
(dictated by eq.~(\ref{sys:evoDM}))
but we just need to follow how any initial energy decreases while crossing the Earth.
This process is well approximated by the following system of two ordinary differential equations:
\begin{eqnsystem}{sys:evo}
 \frac{dN_0}{d\ell}  &=& -N_N\sigma_{\rm CC} N_0+ \frac{1-N_0}{1/\tau + 1/N_N\sigma_{\rm CC}}\\
 \frac{dE}{d\ell} &=& N_0 \frac{dE_0}{d\ell} + (1-N_0) \frac{dE_\pm}{d\ell}.
 \end{eqnsystem}
 where 
  $0\le N_0\le 1$ is the fraction of $\DM$ present in neutral state,
 and $1-N_0$ is the fraction of $\DM$ present in charged state.
 The initial conditions are $N_0(0)=1$ and $E(0) = E_{\rm in}$.


\medskip

\begin{figure}[t]
\begin{center}
\includegraphics[width=0.48\textwidth]{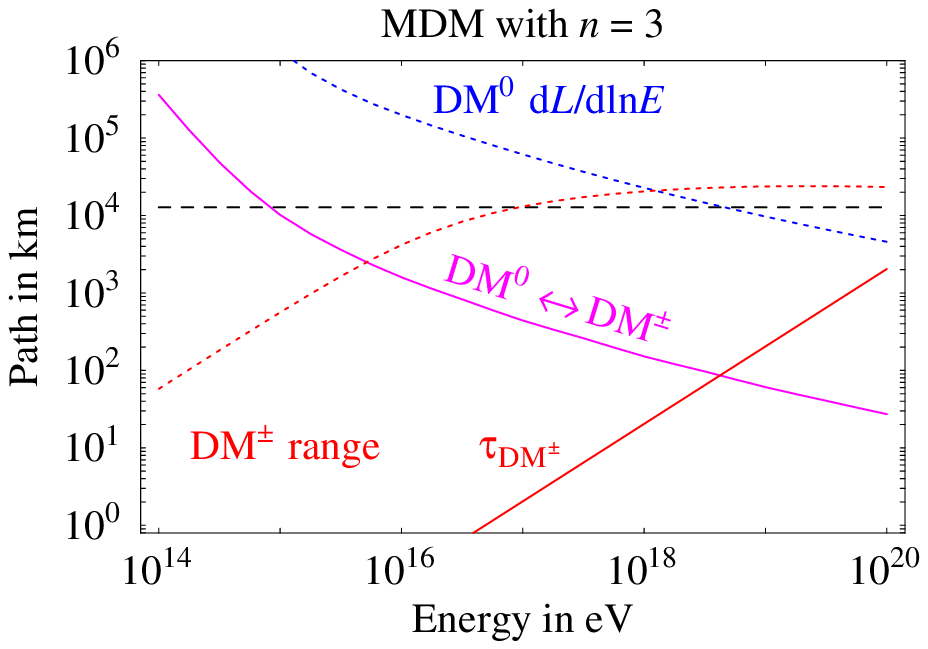}
\includegraphics[width=0.48\textwidth]{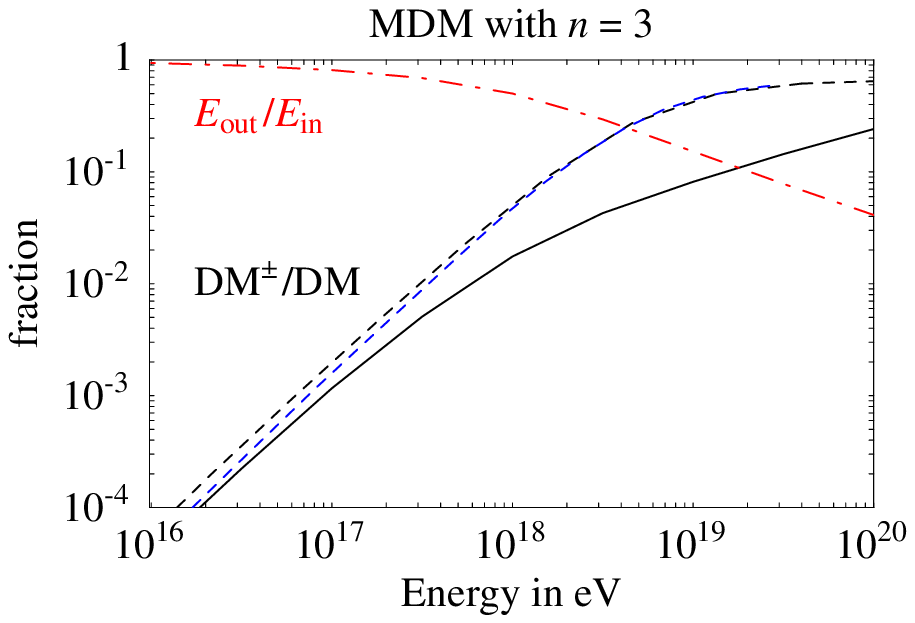}
\caption{\label{fig:range}\em 
We consider the wino-like MDM: a fermion with $n=3$ and 
$M=2.7\TeV$.
{\bf Left plot:} 
The dotted lines show the length scales that characterize energy losses in matter with $\rho = 6.3\,{\rm g/cm}^3$.
More precisely: with the upper blue dotted line we show the effective 
$1/\beta_{\rm CC}$ experienced by a $\DM^0$ as if it never transformed in the charged state, and with the red dotted line the range of a stable $\DM^\pm$.
The solid lines show the effects of neutral to charge transformations and vice-versa: the solid ascending red line is the $\DM^\pm$ life-time while the solid descending pink line is the mean free path for $\DM^0\to \DM^\pm$ CC interactions.
{\bf Right plot:} we consider a $\DM^0$ which crosses the Earth vertically with
initial energy $E_{\rm in}$ and we show 
$E_{\rm out}/E_{\rm in}$ (upper red dot-dashed line) and
the fraction of $\DM^\pm$ over the total number of DM particles (\,$\DM^0 + \DM^\pm$). These fractions are shown as a function of $E_{\rm out}$ in the simplified numerical approach (black dashed line) and in the full numerical approach (blue dashed line, almost superimposed to the former) as well as a function of $E_{\rm in}$ (black solid line). 
}
\end{center}
\end{figure}

Fig.\fig{range}a shows the numerical relevance of the various effects,
and allows to understand their interplay.
The $\DM^0$ interaction length starts to be smaller than the Earth thickness at $E\circa{>}10^{15}\eV$, 
but these interactions have no effects, since the produced $\DM^\pm$ regenerates $\DM^0$ in a negligible length
and with negligible energy losses.
At $E\circa{>}10^{18}\,{\rm eV}$ energy losses start to be moderately relevant,
and at roughly the same energy the $\DM^{0,\pm}$ mean free-path in Earth matter 
becomes comparable to the $\DM^\pm$ decay length:
the Earth is crossed loosing an order one fraction of the initial energy,
and spending an order one fraction of the path as $\DM^\pm$ rather than as $\DM^0$.

\bigskip

Fig.\fig{range}b shows the numerical results: the black continuous line
shows the $\DM^\pm$ fraction $1-N_0$ as function of the initial energy,
after vertically crossing the Earth.
The dashed line shows the same fraction as function of the final energy.
Actually, we have two dashed lines obtained by solving the full and simplified
system of evolution equations:
their agreement confirms the validity of the continuum energy loss approximation.
(In the full case we assumed an initial $\DM^0$ energy spectrum proportional to the
CR energy spectrum). 
These $\DM^\pm$ fraction is roughly given by $1-N_0 \approx (1+N_N\sigma_{\rm CC}/\tau)^{-1}$, and
depends on the local nucleon density $N_N$,
averaged over a typical scale of tens of km.
This figure allows to compute the $\DM^\pm$ rate for any initial energy spectrum of $\DM$.
The upper red dot-dashed line shows the ratio $E_{\rm out}/E_{\rm in}$ between the
final and initial energy.

\begin{figure}[t]
\begin{center}
\includegraphics[width=0.48\textwidth]{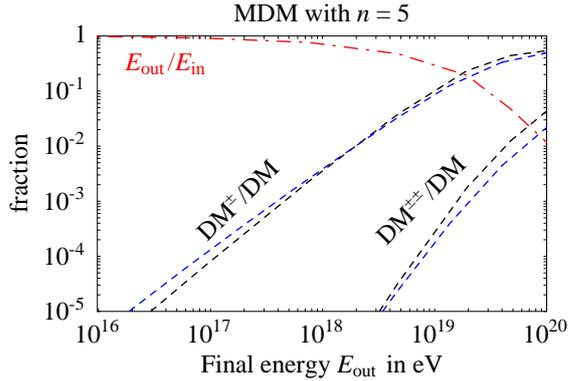}
\caption{\label{fig:range5}\em 
We consider the MDM fermion with $n=5$ and 
$M=10\TeV$ which crosses the earth vertically with
initial energy $E_{\rm in}$ and final energy $E_{\rm out}$. We show 
$E_{\rm out}/E_{\rm in}$ (upper red dot-dashed line) and
the fraction of $\DM^\pm$ and of $\DM^{\pm\pm}$
over the total number of $\DM$ particles as function of $E_{\rm out}$.}
\end{center}
\end{figure}

Fig.\fig{range5} shows the analogous result for the fermion 5-plet, which
has a $\DM^{\pm\pm}$ component with a much shorter life-time $\approx 0.05\,{\rm cm}$.
Due to its fast decay,
the new $\DM^{\pm\pm}$ component is not present
with a significant abundance at the UHE CR energies $E\circa{<} 10^{20}\eV$.

\subsection{Production of UHE Minimal Dark Matter}\label{UHEprod}
Minimal DM particles may exist at ultra high energies if they are a component of the primary cosmic rays, if they manage to be produced by standard UHE CR in the earth's atmosphere or if more exotic CR scenarios will prove to be motivated. Let us look at these opportunities in turn.

Although the mechanism that accelerates the UHE primary CR has not yet been established,
the plausible astrophysical standard mechanism (first-order Fermi acceleration in magnetic shock waves around various objects, such as gamma-ray bursts and
supernova remnants) accelerates stable charged particles.
In the MDM scenario, the neutral $\DM^0$ is accompanied by a charged
partner with life-time $\tau \sim{\rm cm}$:
this value is not macroscopic enough to lead to acceleration of $\DM^\pm$ particles.
Moreover, the standard mechanism accelerates protons or charged nuclei in regions that are
believed to be transparent: i.e.\ the local density is so small
these primary CR particles negligibly hit on the surrounding DM accelerating it,
or on the surrounding material producing DM pairs.

Production of DM pairs might be relevant when UHE CR particles hit the Earth.
The number of DM particles generated by one
UHE proton with energy $E\circa{>} (2M)^2/m_p \sim 10^{18}\eV$
(which is the energy range that leads to a sizable $\DM^\pm$ fraction)
is $r\sim \sigma/\sigma_{pp}\sim 10^{-11}$,
where $\sigma_{pp}\sim 1/m_\pi^2$ is the total hadronic cross section,
and $\sigma$ is the $pp\to \DM\,\DM$ cross section, computed in~\cite{MDM}.
A UHE electron neutrino produces a higher fraction of DM, 
$r\sim \sigma/\sigma_{\nu} \sim 10^{-6}$, 
but only at higher energies $E\circa{>} (2M)^2/m_e$.
These processes lead to a double DM signature, but their rate looks too small for present detectors
(see e.g.~\cite{neutralinotransport,Reno}).

A UHE DM rate detectable in forthcoming experiments can arise in more speculative scenarios: e.g.\ they may be a component of the UHE CR generated by decays of ultra-heavy particle, in the so called `top-down' scenarios~\cite{UHP}.

In terms of absolute numbers, 
if DM constitutes a number fraction $r$ of CR with energy above $10^{17}\eV$ ($10^{18}\eV$),
one expects a flux of $\approx 60r$
($\approx 6r$) $\DM^\pm$ per km$^2\cdot{\rm yr}$, in the case $n=3$.
Atmospheric neutrinos generate a flux of $\approx 10^3$ ($\approx 10$)
up-going muons per km$^2\cdot{\rm yr}$ with $E>10^{13}\eV$ $(10^{14}\eV$):
as discussed in the next section, this is a background to $\DM^\pm$ searches.

\subsection{Detection of UHE $\DM^\pm$}
\label{UHEdetection}

Let us study how UHE $\DM^\pm$ can be searched for.
In a detector like IceCUBE or Antares $\DM^\pm$ roughly look like muons
with fake energy $E_\mu = E_\pm \beta_{\rm CC}/\beta_\mu \sim 10^{-4} E_\pm$,
because muon energy losses are approximatively given by
$-dE_\mu/d\ell = \alpha+\beta_\mu E_\mu$ with
$\beta_\mu \approx 0.2/{\rm kmwe}$~\cite{PDG}:
indeed $\beta$ is roughly inversely proportional to the particle mass.

One therefore needs to carefully study charged tracks to
see the difference between a  $\DM^\pm$ with $E_\pm\sim 10^{18}\eV$
and a muon with $E_\mu\sim 10^{14}\eV$.
The muon would loose all its energy in about $2 \,{\rm kmwe}$
(with a characteristic energy loss profile)
while $\DM^\pm$ have a much longer range in matter.
On the contrary, the muon is essentially stable,
while $\DM^\pm$  decays in about 10 km (less at lower energy: see fig.\fig{range}a), 
suddenly disappearing into a $\DM^0$ and a $\pi^\pm$ with energy $E_\pi \sim m_\pi E_{\pm}/M$.
Furthermore, the  characteristic scale for the $\DM^0\leftrightarrow\DM^\pm$ transformation
is about tens of km.
Unfortunately it seems very difficult to exploit these differences
to discriminate between a $\mu^\pm$ and
a $\DM^\pm$
in the forthcoming IceCUBE  km-size detector~\cite{ICECUBE}, 
because it only has a $1\,{\rm km}$ size and modest granularity.


\begin{figure}[t]
$$\hspace{-5mm}
\includegraphics[width=0.6\textwidth]{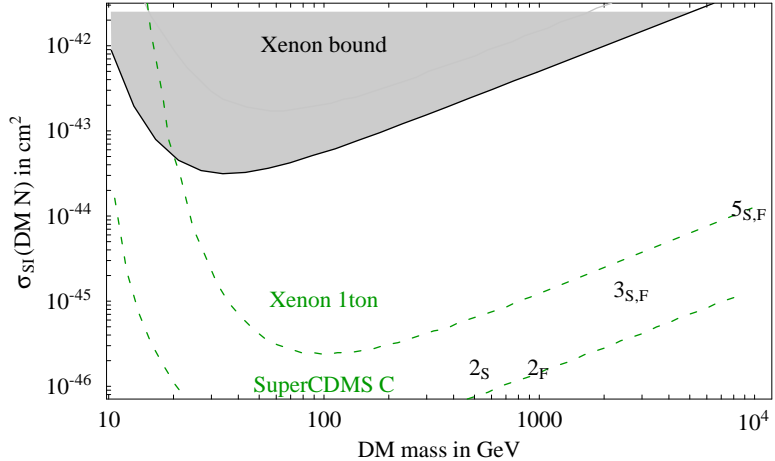}$$
\caption{\label{fig:direct}\em
Predicted mass and predicted spin-independent cross sections per nucleon of MDM candidates.
$S$calar ($F$ermionic) {\rm SU(2)} $n$-tuplets are denoted as $n_S$ ($n_F$).
The shaded region is excluded by~\cite{Xenon} and
the dashed lines indicate the planned sensitivity of future experiments~\cite{future}.
The prediction for doublets with $Y\neq0$ hypercharge holds up to the caveats discussed in the text;
furthermore we assumed the nuclear matrix element $f=1/3$ and the Higgs mass $m_h=115\GeV$.
}
\end{figure}

\section{Conclusions}\label{concl}
We considered DM that fills one SU(2) multiplet
and only interacts with SM vectors (`Minimal Dark Matter'~\cite{MDM}).
This model has one free parameter, the DM mass $M$, which is
fixed from the observed DM abundance assuming that DM is
a thermal relic.
We have reconsidered this computation including
non-perturbative corrections due to Coulomb-like forces
mediated by SM vectors,
that significantly increase $M$.
For example, a fermion 5-plet automatically behaves as MDM
(and in particular is automatically stable): 
its mass increases from $M=4.4\TeV$~\cite{MDM} to $10\TeV$. 
The cosmological freeze-out abundance is plotted as function of $M$ in fig.\fig{F50}a (fermion 5-plet),\fig{F30}a (wino-like fermion 3-plet, $M\approx 2.7\TeV$~\cite{HisanoCosmo}),\fig{S30}a (scalar 3-plet, $M\approx 2.5\TeV$)
and\fig{F2}a (Higgsino-like fermion 2-plet, $M\approx 1.0\TeV$).
We also discussed the scalar 5-plet 
($M\approx 9.4\TeV$) and 7-plet ($M\approx 25\TeV$).

\medskip

Having fixed $M$, we studied the MDM  signals.

\medskip


Concerning the signals at {\em colliders} (see~\cite{MDM,Gunion2}), we here point out
that the charged components $\DM^\pm$ in the MDM multiplet
manifest as non-relativistic cm-scale ionizing charged tracks
negligibly bent by the $B\sim$ Tesla magnetic field present around the interaction region.
$\DM^\pm$ decays with $97.7\%$ branching ratio into $\pi^\pm$,
giving a relativistic track bent by the magnetic fields.
Everything happens in the inner portion of the detector and
this signature is free from SM backgrounds.
However, at a hadron collider like LHC
it seems not possible to trigger on this signature:
level-1 triggers are located  far from the interaction point and
MDM candidates are so heavy that  the event rate is too low.
Unlike in other scenarios we cannot choose a favorable benchmark point
in a vast parameter space.

\medskip

Astrophysics offers more promising detection prospects.

\medskip

The cross section for {\em direct DM detection} negligibly depends on $M$, and remains
the same as in~\cite{MDM}: fig.\fig{direct} summarizes the situation.
Since MDM is much heavier than a typical nucleus,
 direct DM searches can precisely reconstruct
only  the combination $\sigma_{\rm SI}/M$.

Next,  we considered {\em indirect DM detection}, computing how
non-perturbative effects enhance the 
$\DM^0 \DM^0$ annihilations into $W^+W^-$, $\gamma\gamma$, $\gamma Z$, $ZZ$.
Results are shown in fig.\fig{F30}b (wino-like fermion 3-plet~\cite{Hisano}),\fig{S30}b 
(scalar 3-plet),\fig{F50}b (5-plet), and\fig{F2}b (Higgsino-like fermion 2-plet~\cite{Hisano}).
Signal rates can now be computed by
combining this particle physics with (uncertain) astrophysics:
for example fig.\fig{gamma} shows the predicted spectrum  of galactic $\gamma$.

\smallskip

Finally, we assumed that some DM particles are present among the cosmic rays at Ultra High Energies
and identified an unusual signal, characteristic of heavy and quasi-degenerate multiplets
containing neutral and charged components:
a $\DM^0$ crosses the Earth without loosing much energy and,
at $E\circa{>}10^{17}\eV$,
a sizable fraction of the travel is done as $\DM^\pm$,
leading to charged tracks, that might be detectable in neutrino telescopes such as IceCUBE
or km3net~\cite{Antares}.
Fig.s\fig{range} and \fig{range5} quantitatively show how frequent this phenomenon is.

\paragraph{Acknowledgements}
We thank Masato Senami for several clarifications about~\cite{HisanoCosmo}. We also thank Gianfranco Bertone, J\"urgen Brunner, Giacomo Cacciapaglia, Paschal Coyle, Michele Frigerio, Dario Grasso, John Gunion, Teresa Montaruli, Slava Rychkov, G\"unter Sigl, Igor Sokalski and Igor Tkachev for useful discussions. 
The work of M.C. is supported in part by INFN under the postdoctoral grant 11067/05 and in part by CEA/Saclay. 

\appendix

\section{Annihilation cross sections}\label{Pert}
We here give results for the tree-level DM DM annihilation cross sections.
We define the adimensional `reduced cross section'
\beq \hat\sigma(s)= \int_{-s}^0 dt
\sum \frac{|\mathscr{A}|^2}{8\pi s}
\eeq
where $s,t$ are the Madelstam variables and the sum runs over
all DM components and over all SM vectors, fermions and scalars,
assuming that all SM masses are negligibly small.

 The DM abundance is computed by solving the Boltzmann equation
 \beq \label{eq:Bol} sZHz \frac{dY}{dz} =
  -2\bigg(\frac{Y^2}{Y^{2}_{\rm eq}}-1\bigg)\gamma_A,\qquad
  \gamma_A =\frac{T}{64 \pi^4} \int_{4M^2 }^{\infty} ds~ s^{1/2}
 {\rm K}_1\bigg(\frac{\sqrt{s}}{T}\bigg) \hat{\sigma}_A(s)\eeq
  where $z=M/T$, K$_1$ is a Bessel function, $Z=( 1 - \frac{1}{3} \frac{z}{g_s}\frac{d g_s}{dz})^{-1}$,
  the entropy density of SM particles is $s={2\pi^2}g_{*s}T^3/45$,
  $Y=n_{\rm DM}/s$ where $n_{\rm DM}$ is the number density of DM particles plus anti-particles,
  and $Y_{\rm eq}$ is the value that $Y$ would have in thermal equilibrium.
We can write a single equation for the total DM density because DM scatterings with
SM particles maintain thermal equilibrium within and between the single components.
  We ignored the Bose-Einstein and Fermi-Dirac factors as they are negligible at the
  temperature $T \sim M/26$ relevant for DM freeze-out.
  
We assume that DM fills one SU(2) multiplet with dimension $n$ and
hypercharge $Y$; when $Y\neq 0$ the conjugate multiplet is also added
in order to allow for a gauge-invariant DM mass term.
For example, the Higgsino has $n=2$ and $Y=1/2$; the wino has $n=3$ and $Y=0$.
We define $g_{\cal X}$ as the number of DM degrees of freedom:
$g_{\cal X}=n$ for scalar DM with $Y=0$;
$g_{\cal X}=2n$ for scalar DM with $Y\neq 0$ and for spin-1/2 (Majorana) DM with $Y=0$;
$g_{\cal X}=4n$ for spin-1/2 (Dirac) DM with $Y\neq 0$.
For fermion DM we get
\begin{eqnarray}\nonumber
\hat\sigma_A &=& \frac{g_{\cal X}}{24 \pi  n}\nonumber
 \left[(9C_2-21C_1)\beta+(11C_1-5C_2)\beta^3-3 \bigg(2 C_1 (\beta^2-2)+C_2(\beta^2-1)^2\bigg) \ln\frac{1+\beta}{1-\beta }\right]\\
 &&+ g_{\cal X} \bigg(\frac{3g_2^4(n^2-1)+20 g_Y^4 Y^2}{16\pi}+\frac{g_2^4 (n^2-1)+4g_Y^4 Y^2}{128\pi}\bigg)
\bigg(\beta - \frac{\beta^3}{3}\bigg)
\end{eqnarray}
and for scalar DM we get
\begin{eqnarray}
\hat\sigma_A &=&  \frac{g_{\cal X}}{24 \pi  n}\nonumber
 \left[(15 C_1 - 3 C_2) \beta + (5C_2 - 11 C_1) \beta^3+3(\beta^2-1)\bigg(2 C_1+C_2(\beta^2-1)
 \bigg) \ln\frac{1+\beta}{1-\beta }\right]\\
&&+ g_{\cal X} \bigg(\frac{3g_2^4 (n^2-1)+20 g_Y^4 Y^2}{48 \pi}+\frac{g_2^4 (n^2-1)+4g_Y^4 Y^2}{384\pi}\bigg)\cdot \beta^3
\end{eqnarray}
where $x=s/M^2$ and $\beta = \sqrt{1-4/x}$ is the DM velocity in the DM DM center-of-mass frame.
The first line gives the contribution of annihilation into vectors,
the second line contains the sum of the contributions of annihilations into
SM fermions and vectors respectively.
The gauge group factors are defined as
\begin{eqnarray}
C_1 &=&\sum_{A,B}  {\rm Tr}\,  T^A T^A T^B T^B =
g_Y^4 nY^4 + g_2^2 g_Y^2 Y^2 \frac{n(n^2-1)}{2}+g_2^4 \frac{n(n^2-1)^2}{16}\\
C_2 &=& \sum_{A,B}  {\rm Tr}\, T^A T^B T^A T^B =  \noindent
g_Y^4 nY^4 + g_2^2 g_Y^2 Y^2 \frac{n(n^2-1)}{2}+g_2^4 \frac{n(n^2-1)(n^2-5)}{16}
\end{eqnarray}
where the sum is over all SM vectors $A=\{Y,W^1,W^2,W^3\}$ with gauge coupling
generators $T^A$.

\medskip

The DM freeze-out abundance is accurately determined by the leading
two terms of the expansion of for small $\beta$, that describe
the $s$-wave and the $p$-wave contributions.
This approximation allows to analytically do the thermal average in eq.\eq{Bol}:
\beq
\hat\sigma_A \stackrel{\beta\to 0}{\simeq} c_s \beta + c_p \beta^3 +\cdots\qquad\hbox{implies}\qquad
 \gamma_A \stackrel{\beta\to 0}{\simeq} 
\frac{MT^3e^{-2M/T}}{32\pi^3}\left[c_s + \frac{3T}{2M}(c_p+\frac{c_s}{2})+\cdots\right].
\eeq
More complex models of DM can have extra interactions beyond gauge interactions,
but the gauge contribution we computed accurately is the minimal model-independent contribution present
in all cases.
Eq.~\eq{Bol} also arises  as a subset of the Boltzmann equations that describe
leptogenesis from decays of gauge-interacting particles:
the scalar triplet corresponds to $n=3$, $Y=1$~\cite{scalartriplet},
and the fermion triplet to $n=3$ and $Y=0$~\cite{fermiontriplet}.

\footnotesize

\begin{multicols}{2}
  
\end{multicols}


\begin{thebibliography}{nn}
\bibitem{Zcoupling}
The upper bound is imposed by the null results of direct detection searches, see~\cite{Xenon}. See also \hepart[0706.0526]{G. Belanger, A. Pukhov and G. Servant} for a recent analysis of this feature as a phenomenological assumption and in specific models.
\end{thebibliography}
\end{document}